\documentclass{svproc}
\usepackage{url}

\makeatletter
\renewcommand\footnoterule{%
  \kern-3\p@
  \hrule\@width.4\columnwidth
  \kern2.6\p@}
  \makeatother
\usepackage[dvipsnames]{xcolor}
\usepackage{cite}
\usepackage{hyperref}

\usepackage{amsmath,amssymb,amsfonts}
\usepackage{algorithmic}
\usepackage{graphicx}
\usepackage{textcomp}
\usepackage{centernot}
\usepackage[braket, qm]{qcircuit}
\usepackage{braket}
\def\BibTeX{{\rm B\kern-.05em{\sc i\kern-.025em b}\kern-.08em
    T\kern-.1667em\lower.7ex\hbox{E}\kern-.125emX}}
\usepackage{multicol}
\usepackage{subcaption}
\usepackage{bbding}
\usepackage{wrapfig}
\usepackage{tabularx}
\hypersetup{
    colorlinks,
    linkcolor={black},
    citecolor={black},
    urlcolor={blue}
}
\usepackage{tikz}
\usepackage{pgfplots} 
\usetikzlibrary{pgfplots.groupplots}
\usetikzlibrary{intersections}
\pgfplotsset{compat=1.16}
\pgfplotsset{scaled y ticks=false}
\usepgfplotslibrary{fillbetween}
\usepackage{mathtools}
\usepackage{listings}
\usepackage{url}
\usepackage{booktabs}

\usepackage{array}
\newcommand{\PreserveBackslash}[1]{\let\temp=\\#1\let\\=\temp}
\newcolumntype{C}[1]{>{\PreserveBackslash\centering}p{#1}}
\tikzset{SubCaption/.style={
text width=2in,yshift=-3mm, align=center,anchor=north, 
}}
\providecommand{\keywords}[1]{\textbf{\small{Keywords --}} #1}

\usepackage[none]{hyphenat}
\usepackage{flushend}
\usepackage{algorithm2e}
\RestyleAlgo{ruled}

\begin{document}
\sloppy
\mainmatter              

\title{Quantum Computing Techniques for Multi-Knapsack Problems}
\titlerunning{\textit{QC for Multi-Knapsack Problems}}
\author{Abhishek Awasthi\inst{1}\and
Francesco B\"ar\inst{2}\and 
Joseph Doetsch\inst{3}\and
Hans Ehm\inst{4}\and 
Marvin Erdmann\inst{5}\and 
Maximilian Hess\inst{4}  \and 
Johannes Klepsch\inst{5} \and 
Peter A. Limacher\inst{2}  \and 
Andre Luckow\inst{5}  \and 
Christoph Niedermeier\inst{6} \and 
Lilly Palackal\inst{4} \and 
Ruben Pfeiffer\inst{4} \and 
Philipp Ross\inst{5} \and 
Hila Safi\inst{6}  \and 
Janik Sch\"onmeier-Kromer\inst{2} \and 
Oliver von Sicard\inst{6} \and 
Yannick Wenger\inst{3} \and 
Karen Wintersperger\inst{6} \Envelope \and 
Sheir Yarkoni\inst{7}}
\authorrunning{\textit{Quantum Technology and Application Consortium – QUTAC}} %

\institute{
BASF SE, Ludwigshafen am Rhein, Germany, \\
\and
SAP SE, Walldorf, Germany, \\
\and
Lufthansa Industry Solutions AS GmbH, Raunheim, Germany, \\
\and
Infineon Technologies AG, Neubiberg, Germany, \\
\and
BMW AG, Munich, Germany, \\
\and
Siemens AG, Munich, Germany, \\
\and
Volkswagen AG, Munich, Germany. \\ 
\textit{This paper was developed within the Quantum Technology and Application Consortium (QUTAC).}\thanks{{\Envelope } K. Wintersperger: karen.wintersperger@siemens.com, QUTAC: info@qutac.de \\ \\
\scriptsize{This is an Author Accepted Manuscript version of the following chapter: Awasthi et al., Quantum Computing Techniques for Multi-Knapsack Problems, published in Intelligent Computing, edited by Kohei Arai, 2023, Springer Cham reproduced with permission of Springer Cham. The final authenticated version is available online at: \url{https://doi.org/10.1007/978-3-031-37963-5_19}}
}
}


\maketitle

\begin{abstract}	
Optimization problems are ubiquitous in various industrial settings, and multi-knapsack optimization is one recurrent task faced daily by several industries. The advent of quantum computing has opened a new paradigm for computationally intensive tasks, with promises of delivering better and faster solutions for specific classes of problems. This work presents a comprehensive study of quantum computing approaches for multi-knapsack problems, by investigating some of the most prominent and state-of-the-art quantum algorithms using different quantum software and hardware tools. The performance of the quantum approaches is compared for varying hyperparameters. We consider several gate-based quantum algorithms, such as QAOA and VQE, as well as quantum annealing, and present an exhaustive study of the solutions and the estimation of runtimes. Additionally, we analyze the impact of warm-starting QAOA to understand the reasons for the better performance of this approach. We discuss the implications of our results in view of utilizing quantum optimization for industrial applications in the future. In addition to the high demand for better quantum hardware, our results also emphasize the necessity of more and better quantum optimization algorithms, especially for multi-knapsack problems.
\newline \\
\end{abstract}

\section{Introduction}
The knapsack problem deals with a set of items, each having a value and a weight, which are assigned to a knapsack with a certain capacity. The task is to maximize the total value of items placed in the knapsack while respecting the capacity of the knapsack. This optimization problem is also referred to as a one-dimensional 0/1 knapsack problem and is known to be NP-hard \cite{Karp1972}. Generalizing the problem to $M$ knapsacks (multi-knapsack problem) and valuing items differently in each knapsack complicates the problem further \cite{pisinger}. However, these more delicate versions of the knapsack problem can be used as models for many real-world use cases. The scale of many industrial applications, combined with the multi-objective nature of the $M$-dimensional 0/1 knapsack problem, poses a significant challenge to classical solution approaches~\cite{pisinger}. Therefore, new computational paradigms such as quantum computing are explored, with the goal of finding a speed-up over traditional approaches~\cite{Ebadi_2022, Streif_2021}. Quantum computing allows solving certain types of complex problems significantly faster than classical devices~\cite{grover, shor}. Due to its broad applicability, the knapsack problem and its variants have been well studied, and several classical approaches have been proposed to find solutions to this class of problems. Surveys on heuristic algorithms for multiple categories of knapsack problems can be found in \cite{laabadi} and \cite{wilbaut}.

The field of research dedicated to quantum computing solutions for knapsack problems is considerably younger and smaller. Two techniques that are built upon the quantum approximate optimization algorithm (QAOA) are introduced in \cite{vanDam}: one of them ``warm-starts'' the quantum optimization algorithm by seeding it with an initial solution using a greedy classical method, and the other uses special mixing Hamiltonians to improve the exploration of the solution space. Results indicate that both approaches outperform similarly shallow classical heuristics in one-dimensional knapsack problem instances. Reformulations of this version of the knapsack problem are evaluated in~\cite{quintero}. For the multi-knapsack problem, two quantum-inspired evolutionary algorithms (QIEA) are presented in \cite{QIEA}, solving knapsack problem instances with more than 10,000 items. Besides these promising findings of the potential of quantum computing for the knapsack use case, there are studies indicating challenges and even an inability of quantum optimization techniques to outperform classical methods, at least in the era of near-term noisy intermediate-scale quantum (NISQ) devices. In \cite{pusey-nazarro}, it is shown that a \emph{D-Wave 2000Q} quantum annealer could not provide optimal solutions for many small-scale knapsack problems due to the limitations of the hardware. A general theory on the limitations of optimization algorithms on NISQ devices is presented in~\cite{Franca}. Since most of the optimization algorithms for NP-hard problems are heuristic in nature, frameworks such as QUARK~\cite{quark} are essential to obtain a thorough comparison of the performances of classical and quantum algorithms on various hardware backends.

This work presents a comprehensive benchmark of different quantum algorithms for a use case relevant to many industries. We study an $M$-dimensional 0/1 knapsack problem and carry out an extensive comparison of results obtained via QAOA, warm-start QAOA, VQE, Quantum Annealing as well as the iterative heuristic solver with Simulated Annealing. Much work about benchmarking quantum algorithms for optimization problems focuses mainly on a single type of algorithm (circuit model or adiabatic model). However, to identify the most promising quantum algorithm, holistic benchmarking of several quantum approaches needs to be carried out. We study the suitability of quantum algorithms for the multi-knapsack problem with the help of key performance indicators (KPIs) relevant to industrial applications. Since we are interested in the performance of the algorithms, we benchmark the gate-based algorithms on noiseless simulators (while quantum annealing has been carried out on quantum hardware). Nonetheless, in contrast to most of the studies published in this field, we present a practical estimation of the runtimes on quantum hardware for the QAOA and VQE algorithms.
Additionally, we compare the warm-start QAOA described in \cite{Egger_2021} as well as another new variant of warm-start QAOA with the standard algorithms mentioned above, which is another novel contribution of this work to the best of our knowledge.

The remainder of the paper is structured as follows: Section~\ref{sec:business_motivation} provides the business motivation and possible use cases. The modeling of the multi-knapsack problem is described in Section~\ref{modeling}, while Section~\ref{algorithms} gives a brief overview of the quantum algorithms being used, \emph{i.e.}, QAOA, VQE and Quantum Annealing. The information about the problem instances and the definitions of the KPIs used in this work are provided in Section~\ref{results}. We then present our benchmarking results for the studied algorithms, along with the runtimes using a quantum annealer and an estimation of the runtimes of the gate-based algorithms on a real quantum device. We conclude our work with Section~\ref{conclusion} and provide an analysis of the results obtained as well as an outlook. The complete python code for our implementations is available at \url{https://github.com/QutacQuantum/Knapsack}.

\section{Business Motivation}~\label{sec:business_motivation}
The Quantum Technology and Application Consortium (QUTAC) and its members focus on industry use cases for quantum computing applications~\cite{QUTAC}. The general knapsack problem has many applications in decision-making processes along the entire value chain, such as optimizing portfolios~\cite{Kellerer2004}, business operations and supply chains. A prominent example of a multi-knapsack problem in many industries, including the automotive, semiconductor and chemical industry, is the optimization of complex supply chains, since products are usually processed in a global manufacturing network rather than in a single factory. Hence, the need for planning and communication between the manufacturing sites emerges to realize an optimal global manufacturing process. Optimization techniques are crucial to addressing common supply chain challenges and increasing the responsiveness to disruptions, e.g., optimizing freight and warehouse capacities, labor planning and carbon emissions, thereby also enhancing the overall sustainability~\cite{mck_supplychain}. For the semiconductor industry alone, where supply chain management is particularly complex and dynamic, optimization plays a major role in the context of integrated supply chain planning, daily. One major goal is to solve the demand-capacity matching, which corresponds to maximizing the available product units relative to the units promised to the customer. This task consists of several computationally hard optimization problems, since more than a million order confirmations and reorder confirmations have to be regularly calculated. Therefore, high-quality solutions need to be delivered in a reasonable time. In the following, we will restrict ourselves to a simplified model in which we assume that the demand, capacity, and costs are given. Note that this is a strong assumption, as determining the inputs of the model is a computationally hard task itself. The simplified demand-capacity match can be encoded as a multi-knapsack problem.

\section{Modeling}~\label{modeling}
In this section, we present a formal description of the multi-knapsack optimization problem, along with the mathematical formulation of the QUBO model. Given $N$ items and $M$ knapsacks, the objective is to assign as many valuable items as possible to each of the knapsacks while not exceeding the capacity of any knapsack. The problem can be stated as follows. Let $j \in \{0,1,\dots, N-1\}$, then $w_j \in \mathbb{N}_0$ denotes the weight of item $j$ and $v_{i,j}\in \mathbb{N}_0$ denotes the value of item $j$ in knapsack $i \in \{0,1,\dots,M-1\}$. The capacity of a knapsack $i$ is denoted by $c_i \in \mathbb{N}_0$. We define a decision variable $x_{i,j}$, such that $x_{i,j}=1$ if and only if item $j$ is assigned to knapsack $i$, and $0$ otherwise. We can now formulate the corresponding QUBO model, including the problem constraints and objective term.
\begin{itemize}
\item Any item $j$ can be assigned to at most one knapsack. It is possible that an item is not assigned to any knapsack.
\begin{equation}
\small
H_{\text{single}}= \sum_{j=0}^{N-1} \left(\sum_{i=0}^{M-1} x_{i,j}\right)\cdot \left(\sum_{i=0}^{M-1} x_{i,j}-1 \right) \;.    
\label{eq_2}
\end{equation}
\item Ensure that no knapsack's capacity is exceeded. This is achieved by introducing slack bits $y_{i,b}$ with binary expansion, based on the work of Lucas~\cite{lucas_2014}.
The filling of knapsack $i \in \{0,\dots,M-1\}$, which can be smaller than its capacity, is thereby expressed as $c_i-\sum_{b=0}^{\lfloor\log_2{c_i}\rfloor} 2^{b}\cdot y_{i,b}$. Using this formulation, no fillings larger than the knapsack capacity can be encoded, also when $c_i$ is not a power of two. If the sum corresponds to a number larger than $c_i$, the actual filling becomes negative and thus leads to an even larger penalty in the Hamiltonian. The capacity term of the Hamiltonian reads
\begin{equation}
\small
\begin{split}
    H_{\text{capacity}}=&\sum_{i=0}^{M-1} \left[\left(\sum_{j=0}^{N-1} w_j\cdot x_{i,j}\right) +\left(\sum_{b=0}^{\lfloor\log_2{c_i}\rfloor} 2^{b}\cdot y_{i,b}\right) - c_i \right]^2 . 
\end{split}
\label{eq_3}
\end{equation}

\item The objective term is formulated such that our original maximization objective function is converted to a minimization problem, as shown below.
\begin{equation}
\small
H_{\text{obj}}= - \sum_{i=0}^{M-1} \sum_{j=0}^{N-1} v_{i,j} \cdot x_{i,j} \;.  
\label{eq_5}
\end{equation}
\end{itemize}
With the above QUBO terms, we can now formulate the complete QUBO for multi-knapsack optimization problems as $H$, where 
\begin{equation}
\small
H = A\cdot H_{\text{single}} + B\cdot \text{H}_{\text{capacity}} + C \cdot H_{\text{obj}}.  
\label{eq_H_prefactors}
\end{equation}
The coefficients $A,B > 0$ are the penalty weights, $C > 0$ is the objective weight. The minimization of $H$ results is an optimal solution to the multi-knapsack problem.
We have to choose $A, B, C$ such that $\frac{A}{C} > \max_{i,j}(v_{i,j})$ and $\frac{B}{C} > \max_{i,j}(v_{i,j})$. This ensures that any value gained by breaking a constraint is offset by an even larger penalty. Without loss of generality, we therefore choose $C=1$ and set $A = B = 2\cdot \max_{i,j}(v_{i,j})$.

\section{Quantum Optimization Algorithms}~\label{algorithms}
This work provides a comparison for the multi-knapsack problem between several quantum algorithms. Since most of these algorithms are well-known and explained in detail in the literature, we provide only a short overview below.
\subsection{Quantum Approximate Optimization Algorithm (QAOA)}\label{sub_qaoa}
The QAOA is a popular variational algorithm inspired by the adiabatic theorem, devised to produce approximate solutions for combinatorial optimization problems~\cite{farhi}. For brevity, we outline the basics of the algorithm below; for an in-depth explanation of the algorithm we refer the reader to~\cite{zhou}. The QAOA algorithm optimizes any Hamiltonian $\mathcal{C}$ by constructing a predefined parameterized quantum circuit and optimizing the circuit parameters by utilizing classical iterative algorithms. 

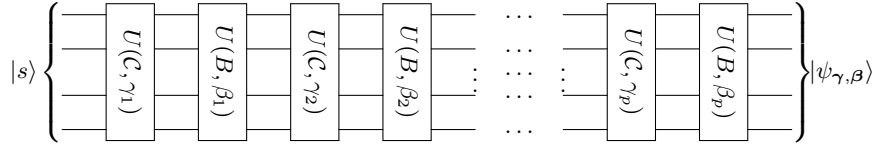
\begin{figure}[ht!]
\centering
$$ 
\Qcircuit @C=1.8em @R=0.5em {
& \multigate{4}{\rotatebox{270}{$U(\mathcal{C},\gamma_1)$}} & \multigate{4}{\rotatebox{270}{$U(B,\beta_1)$}} & \multigate{4}{\rotatebox{270}{$U(\mathcal{C},\gamma_2)$}} & \multigate{4}{\rotatebox{270}{$U(B,\beta_2)$}} & \qw & \hdots & & \multigate{4}{\rotatebox{270}{$U(\mathcal{C},\gamma_p)$}} & \multigate{4}{\rotatebox{270}{$U(B,\beta_p)$}} & \qw \\
& \ghost{\rotatebox{270}{$U(\mathcal{C},\gamma_1)$}} & \ghost{\rotatebox{270}{$U(B,\beta_1)$}} & \ghost{\rotatebox{270}{$U(\mathcal{C},\gamma_2)$}} & \ghost{\rotatebox{270}{$U(B,\beta_2)$}} & \qw & \hdots & & \ghost{\rotatebox{270}{$U(\mathcal{C},\gamma_p)$}} & \ghost{\rotatebox{270}{$U(B,\beta_p)$}} & \qw \\
\lstick{\ket{s}\hspace{0.5em}} & & & & & \vdots & \hdots & \vdots & & & & \hspace{0.5em}\ket{\psi_{\boldsymbol{\gamma},\boldsymbol{\beta}}} \\
& \ghost{\rotatebox{270}{$U(\mathcal{C},\gamma_1)$}} & \ghost{\rotatebox{270}{$U(B,\beta_1)$}} & \ghost{\rotatebox{270}{$U(\mathcal{C},\gamma_2)$}} & \ghost{\rotatebox{270}{$U(B,\beta_2)$}} & \qw & \hdots & & \ghost{\rotatebox{270}{$U(\mathcal{C},\gamma_p)$}} & \ghost{\rotatebox{270}{$U(B,\beta_p)$}} & \qw\\
& \ghost{\rotatebox{270}{$U(\mathcal{C},\gamma_1)$}} & \ghost{\rotatebox{270}{$U(B,\beta_1)$}} & \ghost{\rotatebox{270}{$U(\mathcal{C},\gamma_2)$}} & \ghost{\rotatebox{270}{$U(B,\beta_2)$}} & \qw & \hdots & & \ghost{\rotatebox{270}{$U(\mathcal{C},\gamma_p)$}} & \ghost{\rotatebox{270}{$U(B,\beta_p)$}} & \qw
{\gategroup{1}{11}{5}{11}{0.9em}{\}}}
{\gategroup{1}{1}{5}{1}{0.9em}{\{}}
}
$$
\caption{Schematic diagram of the QAOA circuit with $p$ layers.}
\label{qaoa_circuit}
\end{figure}

Concretely, the QAOA algorithm requires a quantum circuit to sample a quantum state $\ket{\psi_{\boldsymbol{\gamma}, \boldsymbol{\beta}}}=\left(\prod_{l = p}^{1} U(\mathcal{B},\beta_l) \cdot U(\mathcal{C},\gamma_l)\right)\cdot \ket{s}$, where $\ket{s}$ is the uniform superposition state, $U(\mathcal{B},\beta_l)=e^{-i\beta_l \mathcal{B}}$ is the unitary operator resulting from a mixing Hamiltonian and $U(\mathcal{C},\gamma_l)$ is an operator for the problem Hamiltonian $\mathcal{C}$. The mixing Hamiltonian is defined as the sum of Pauli-X ($\sigma^x$) observables acting on all the $n$ qubits, $\mathcal{B}=\sum_{j=1}^{n}\sigma_j^x$. The optimization task is to maximize/minimize $\bra{\psi_{\boldsymbol{\gamma}, \boldsymbol{\beta}}}\mathcal{C}\ket{\psi_{\boldsymbol{\gamma}, \boldsymbol{\beta}}}$, the expectation value of $\ket{\psi_{\boldsymbol{\gamma}, \boldsymbol{\beta}}}$ given the problem Hamiltonian $\mathcal{C}$. For the implementation of QAOA there are a few hyperparameters like the initialization of $\boldsymbol{\gamma}$, $\boldsymbol{\beta}$ and the number of layers $p$ that need to be specified. We provide the analysis of these aspects in Section~\ref{results}. A schematic diagram of the QAOA circuit is provided in Figure~\ref{qaoa_circuit}.

\subsubsection{Warm-start QAOA:}\label{algo_ws_qaoa} 
Warm-start QAOA (WS-QAOA) is a variant of QAOA developed by Egger~\emph{et~al.}~\cite{Egger_2021}. The difference essentially lies in the initial state and the mixing Hamiltonian, which are defined based on the solution of the relaxed QUBO (which admits variables in $[0,1]$ instead of $\{0,1\}$). Suppose there are in total $L$ variables in the original QUBO and $c_l^*$ be the optimal solution value of the $l$th variable in the relaxed QUBO, the initialization of the WS-QAOA circuit is done with $\ket{\phi^*}$, such that
\begin{equation}
\small
\begin{split}
\ket{\phi^*} &= \bigotimes_{l=0}^{L-1} \hat{R}_Y (\theta_l)\ket{0}^{\otimes L}
 = \bigotimes_{l=0}^{L-1} \left(\sqrt{1-c_l^*} \ket{0} + \sqrt{c_l^*}\ket{1} \right)  \;,
\end{split}
\end{equation}
where $\hat{R}_Y (\theta_l)$ is a rotation gate on the $l$th qubit parameterized by angle $\theta_l=2\arcsin\left(\sqrt{c_l^*}\right)$. 
As we can see, this initialization state ensures that the probability of measuring qubit $l$ in state $\ket{1}$ is $c_l^*$. Another difference to the standard QAOA in WS-QAOA is the mixer Hamiltonian. Instead of the Pauli-X Hamiltonian, WS-QAOA utilizes a Hamiltonian whose ground state is $\ket{\phi^*}$ with eigenvalue $-n$. Formally, the mixer Hamiltonian in WS-QAOA is $\hat{H}_M^{ws} = \sum_{l=0}^{L-1}\hat{H}_{M,l}^{ws}$ such that
\begin{equation}
\small
    \hat{H}_{M,l}^{ws}=\begin{pmatrix}
    2 c_l^*-1 & -2 \sqrt{c_l^*\left(1-c_l^*\right)}  \\
    -2 \sqrt{c_l^*\left(1-c_l^*\right)}  & 1-2 c_l^*
\end{pmatrix} \;.
\end{equation}
The mixer operator is $\exp{(-i \beta \hat{H}_{M,l}^{ws})}$, which is implemented in WS-QAOA using single qubit rotation gates $\hat{R}_Y(\theta_l)\hat{R}_Z(-2\beta)\hat{R}_Y(-\theta_l)$. The implementation of the WS-QAOA in this work is identical to the the warm-start QAOA described by Egger~\emph{et~al.}~\cite{Egger_2021}.

\subsubsection{Warm-started Standard QAOA:} \label{algo_ws_standard_qaoa}
As we will see in our results and the study presented by Egger~\emph{et~al.}~\cite{Egger_2021}, the warm-start QAOA definitely performs much better than the standard QAOA. To clearly understand the effect of warm-starting, \emph{i.e.}, initializing the QAOA (WS-QAOA) circuit with the relaxed solution of a QUBO, we propose a variant of WS-QAOA, in that the circuit is initialized with the relaxed QUBO solution, but the mixer Hamiltonian is unchanged to the standard QAOA. We call this approach of QAOA the warm-started standard QAOA, and for brevity we refer to it as \textbf{WS-Init-QAOA}. The WS-Init-QAOA can be formally expressed as a variational quantum circuit which samples $\ket{\psi_{\boldsymbol{\gamma}, \boldsymbol{\beta}}^{\prime}}= \prod_{l = p}^{1} U(\mathcal{B},\beta_l) \cdot U(\mathcal{C},\gamma_l) \cdot \ket{\phi^*}$, where $\ket{\phi^*}$ is the relaxed QUBO solution, explained in Section~\ref{algo_ws_qaoa}. $U(\mathcal{B},\beta_l)=e^{-i\beta_l \mathcal{B}}$ is the parameterized unitary operator resulting from the Pauli-X mixing Hamiltonian and $U(\mathcal{C},\gamma_l)$ is an operator for the problem Hamiltonian $\mathcal{C}$. Note that the only difference to the WS-QAOA is the mixing Hamiltonian.

\subsection{Variational Quantum Eigensolver}
Like QAOA, the variational quantum eigensolver (VQE)~\cite{Peruzzo_2014} consists of a parameterized quantum circuit $U(\theta)$, where $\theta \in [0,2\pi]^n$ are angles for single-qubit rotation gates. Analogous to QAOA, these parameters are tuned to minimize the expectation value $\bra{\psi(\theta)}\mathcal{C}\ket{\psi(\theta)}$, where $\psi(\theta) := U(\theta) \ket{0}$ is the trial state prepared by the circuit $U(\theta)$ and $\mathcal{C}$ is the Hamiltonian encoding the optimization problem.
The VQE offers more freedom in the choice of the circuit. QAOA can be seen as a special case of VQE, namely if we choose a QAOA circuit as $U(\theta)$. A common requirement for the circuit $U(\theta)$ is \textit{hardware efficiency} \cite{Kandala_2017}, meaning that the circuits consist of parameterized one-qubit rotations and two-qubit entangling gates and are kept relatively shallow in order to deal with short coherence times on NISQ devices. The drawback of using generic ansatz circuits is that a larger number of parameters (compared to the QAOA) is needed to guarantee \textit{expressivity}~\cite{Du_2022} of the circuit,~\emph{i.e.}, the circuit's ability to prepare the ground state of any choice of target Hamiltonian. For a full technical review of methods and best practices for VQE, we refer the reader to~\cite{VQE_review}.
~\label{sec:vqe}
\subsection{Quantum Annealing}
Quantum annealing (QA) is a metaheuristic quantum optimization algorithm inspired by Adiabatic Quantum Optimization (AQO) and Adiabatic Quantum Computing (AQC). The algorithm starts by initializing a system of qubits in a simple-to-prepare optimum (known as the \emph{ground state} or the \emph{initial Hamiltonian}) that is slowly evolved to represent a combinatorial optimization problem expressed as an Ising model or QUBO (known as the \emph{final Hamiltonian})~\cite{nishimori}. At the end of this process, the qubits' states represent a possible solution to the combinatorial optimization problem in the final Hamiltonian, with a non-zero probability of being a global optimum (i.e., the ground state of the final Hamiltonian). The specific evolution parameters which govern the path from initial to final Hamiltonian dictate the success of the QA algorithm, specifically the probability of measuring a ground state of the final Hamiltonian. For a full description of the physics and implementation of QA, we refer the reader to~\cite{Hauke2020}; for a review of applications tested using QA, we refer the reader to~\cite{QAapps}.

Quantum annealing (or any QUBO solving method) can be expanded upon via the Iterative Heuristic Solver (IHS), originally developed by Rosenberg~\emph{et al.}~\cite{Rosenberg_2016}. The general idea of this metaheuristic procedure is to split a QUBO problem into several smaller subproblems and solve these iteratively instead of solving the entire problem at once. Suppose we aim to minimize $\mathbf{x}^T Q \mathbf{x}$, where $Q \in \mathbb{R}^{n \times n}$ is the QUBO matrix and $\mathbf{x} \in \{0, 1\}^n$ the binary solution vector. In the basic version of the IHS, we start with a randomly generated initial configuration of $\mathbf{x}$ and repeatedly perform the following steps:
\begin{enumerate}
    \item Randomly choose $k$ variables from $\mathbf{x}$ and fix the other $n-k$ variables.
    \item Optimize over the $k$ chosen variables using the underlying QUBO solver.
    \item Check for an improvement in solution quality over the previous iteration.
\end{enumerate}
If no improvement is detected in the final step for several iterations, we assume \textit{convergence} and output the final state of the vector $\mathbf{x}$ as the optimized solution. It is to be noted that an iterative metaheuristic like this has a significantly larger runtime by design than other more direct approaches to solving a QUBO problem. 
However, the IHS avoids having to embed the full QUBO matrix of a problem, which is potentially large, on a quantum annealer. This increases the size of potentially solvable problems significantly, considering hardware limits, which makes it worthwhile to test the approach alongside our other presented methods.

\section{Results}~\label{results}
In this section, we first discuss different problem instances considered in this work for the multi-knapsack problem, along with the description of different measures to assess the performance of QAOA, VQE, and quantum annealing. Subsequently, we present detailed results obtained via benchmarking these algorithms over different scenarios. It must be noted that in this work we have executed the QAOA and VQE algorithms using state-vector simulations, while quantum annealing was carried out on real quantum devices. Additionally, we present the comparative performance of quantum annealing versus classical heuristic methods.

\subsection{Problem Instances}
\begin{wraptable}[9]{r}{0.5\textwidth}
\vspace*{-3.5em}
\centering
\caption{\small{Benchmark Scenarios}}
\label{tab:scenarios}
\begin{tabular}{c|c|c|c|c} \hline
\textbf{Scenarios}	&	1	&	2	&	3	&	4 \\ \hline
\textbf{Knapsacks}	&	1	&	2	&	2	&	2 \\
\textbf{Items}	&	8	&	4	&	5	&	6 \\
\textbf{Qubits}	&	12	&	14	&	16	&	19 \\
\textbf{Optimal Sol.} ($\boldsymbol{v_{\mathrm{opt}}}$)	&	22	&	12	&	13	&	13 \\ 
\textbf{Prefactors} ($\boldsymbol{A=B}$) & 32 & 10 & 8 & 8 \\ \hline
\end{tabular}
\end{wraptable}
To compare the performance of the various optimization methods and algorithms, $4$ different instances of the knapsack problem are considered, increasing in problem size and complexity, which are listed in Table~\ref{tab:scenarios}. The optimal solution was derived classically for each problem size using 0/1 integer programming. The maximal value of the optimal item distribution and the prefactors $A$ and $B$ of the penalty terms in the Hamiltonian in Equation~\eqref{eq_H_prefactors} are given in the last two rows of Table~\ref{tab:scenarios}. The parameters for the problem instances were initially chosen randomly and partly modified afterwards in order to ensure different levels of complexity.

\subsection{Measures for Solution Quality}~\label{measures}
When solving the knapsack problem with a quantum program, each solvers' output is a probability distribution of quantum states represented as bitstrings, characterized by executing the program several times and taking measurements. 

To inspect the validity of the solutions, the sum of all penalty terms in the QUBO is evaluated. If the total penalty value is equal to zero, the bitstring is considered a valid solution. Note that in this way, bitstrings encoding an item distribution with a total weight smaller than the knapsack capacity ($\sum_{j=0}^{N-1} w_j\cdot x_{i,j} \leq c_i$) are considered invalid if the corresponding $y_i^b$ slack bits are not correct (\emph{i.e.}, $H_{\text{capacity}} > 0$). The best solution $\mathbf{x}_{\mathrm{min}}$ is defined as the bitstring with the lowest energy. This solution is characterized by the total value of the corresponding item distribution $v_{\mathrm{tot}}$, divided by the value of the known optimal solution $v_{\mathrm{opt}}$. This relative value, also known as the approximation ratio, is averaged over several runs $N_{\mathrm{run}}$ for all valid best solutions with the same parameters to obtain the mean closeness to optimum $C_{\mathrm{opt}}(\mathbf{x}_{\mathrm{min}})$, such that
\begin{equation}
\small
    C_{\mathrm{opt}}(\mathbf{x}_{\mathrm{min}}) = \frac{1}{N_{\mathrm{run}}} \sum_{r=1}^{N_{\mathrm{run}}} \frac{v_{\mathrm{tot}}(\mathbf{x}_{\mathrm{min}, r})}{v_{\mathrm{opt}}} \cdot 100 \;.
\label{eq:closeness_to_opt_def}
\end{equation}
In practice, it is useful to find a distribution of items that is not necessarily optimal, but still has a high $C_{\mathrm{opt}}$ value. We consider all valid bitstrings $\mathbf{x}$ within a probability distribution which have a $C_{\mathrm{opt}}$ value above a certain threshold $C_{\mathrm{lim}}$, which we chose to be $90\%$. For each set of parameters, the amplitudes\footnote{The amplitude of a state $\mathbf{x}$ is defined as the square root of its probability in the sampled solution.} $a(\mathbf{x})$ of all $\mathbf{x}$ with $C_{\mathrm{opt}}(\mathbf{x}) \geq C_{\mathrm{lim}}$ are added and the sums are averaged over all runs. Thus, we define the overlap of the sampled solutions with 0.90-opt solution as $\braket{\mathbf{O}_{90}}$, where
\begin{equation}
\small
    \braket{\mathbf{O}_{90}} = \frac{1}{N_{\mathrm{run}}} \sum_{r=1}^{N_{\mathrm{run}}} \sum_{\mathbf{x}|C_{\mathrm{opt}}(\mathbf{x}) \geq C_{\mathrm{lim}}} a_r(\mathbf{x})\;.
\label{eq:overlaps}
\end{equation}
Note that it might be possible that the optimal solution has a high relative value, but the $\braket{\mathbf{O}_{90}}$ remains small, since there are no further solutions with a relative value above $90\%$.

\subsection{Results for QAOA, WS-QAOA and WS-Init-QAOA}~\label{qaoa_results}
In this section, we discuss detailed results of the QAOA, WS-QAOA and WS-Init-QAOA, implemented using Qiskit~\cite{qiskit_2022}. For all results presented here, the quantum circuits were sampled with $n_{\text{samp}} = 10,000$ shots. The classical optimization of $\boldsymbol{\gamma}$ and $\boldsymbol{\beta}$ is carried out with the off-the-shelf optimizer class of SciPy Python library using the sequential least squares programming (SLSQP) algorithm~\cite{scipy_2022}, where the maximal number of iterations was limited to $10,000$.

\begin{figure}[ht!]
\begin{subfigure}{0.32\linewidth}
\hspace{2em}
\begin{tikzpicture}[trim axis left, trim axis right, scale=0.8, every node/.style={scale=0.8}]
\begin{axis}[
	tick label style={/pgf/number format/fixed},scale only axis,
	xmin=11.7, xmax=19.3,
	xtick={12, 14, 16, 19},
	ymin=-0.02, ymax=0.17,
	ytick={0.0, 0.02, 0.04, 0.06, 0.08, 0.10, 0.12, 0.14, 0.16},
	xlabel=Number of Qubits,
	ylabel style={align=center},
	ylabel={Overlap for $0.90$-opt, $\braket{\mathbf{O}_{90}}$},
	ymajorgrids=true,
	grid style=dashed,
	legend pos=north east,
	legend style={nodes={scale=0.9, transform shape}},
	width=1\linewidth,
	legend columns=2,
	]
\addplot [color=RoyalBlue, mark=diamond*, mark size=1pt, mark options=solid, line width=1.2pt, error bars/.cd,y dir=both,y explicit] plot coordinates {
(12, 0.04115155750126699) +- (0.032046697369353026, 0.032046697369353026)
(14, 0.040596426364068) +- (0.036122991401094214, 0.036122991401094214)
(16, 0.00741421356237309) +- (0.008064083162488021, 0.008064083162488021)
(19, 0.001) +- (0.0029999999999999996, 0.0029999999999999996)
};
\addplot [color=WildStrawberry, mark=diamond*, mark size=1pt, mark options=solid, line width=1.2pt, error bars/.cd,y dir=both,y explicit] plot coordinates {
(12, 0.025989461189141572) +- (0.03905857388090292, 0.03905857388090292)
(14, 0.029924074484064457) +- (0.014900370827906047, 0.014900370827906047)
(16, 0.009650281539872869) +- (0.011396389647543899, 0.011396389647543899)
(19, 0.00141421356237309) +- (0.004242640687119271, 0.004242640687119271)
};
\addplot [color=YellowOrange, mark=diamond*, mark size=1pt, mark options=solid, line width=1.2pt, error bars/.cd,y dir=both,y explicit] plot coordinates {
(12, 0.03124935122593806) +- (0.02132607354946444, 0.02132607354946444)
(14, 0.04097622223897832) +- (0.02807838982046551, 0.02807838982046551)
(16, 0.00997469149468815) +- (0.012127409949396411, 0.012127409949396411)
(19, 0.0) +- (0.0, 0.0)
};
\addplot [color=BlueGreen, mark=diamond*, mark size=1pt, mark options=solid, line width=1.2pt, error bars/.cd,y dir=both,y explicit] plot coordinates {
(12, 0.028308110668036478) +- (0.045778182181063894, 0.045778182181063894)
(14, 0.04711675233820504) +- (0.03313208767754477, 0.03313208767754477)
(16, 0.004) +- (0.0066332495807108, 0.0066332495807108)
(19, 0.0) +- (0.0, 0.0)
};
\addplot [color=Periwinkle, mark=diamond*, mark size=1pt, mark options=solid, line width=1.2pt, error bars/.cd,y dir=both,y explicit] plot coordinates {
(12, 0.033309286345939645) +- (0.029193423203971086, 0.029193423203971086)
(14, 0.04813145759875467) +- (0.0476126393121782, 0.0476126393121782)
(16, 0.01187831517751083) +- (0.010871414850260752, 0.010871414850260752)
(19, 0.0) +- (0.0, 0.0)
};
\addplot [color=Bittersweet, mark=diamond*, mark size=1pt, mark options=solid, line width=1.2pt, error bars/.cd,y dir=both,y explicit] plot coordinates {
(12, 0.04041558542931491) +- (0.03515467842555999, 0.03515467842555999)
(14, 0.04403511993070715) +- (0.047366641954179965, 0.047366641954179965)
(16, 0.017360527953183442) +- (0.018276472947745627, 0.018276472947745627)
(19, 0.001) +- (0.003, 0.003)
};
\legend{p=$1$, p=$2$, p=$3$, p=$4$, p=$5$, p=$6$}
\end{axis}
\end{tikzpicture}
\centering
\caption{\small{Standard QAOA.}}
\label{res_qaoa_overlap}
\end{subfigure}%
\hfill
\begin{subfigure}{.32\linewidth}
\hspace{2.1em}
\centering
\begin{tikzpicture}[trim axis left, trim axis right, scale=0.8, every node/.style={scale=0.8}]
\begin{axis}[
tick label style={/pgf/number format/fixed},
scale only axis,
	xmin=11.7, xmax=19.3,
	xtick={12, 14, 16, 19},
	ymin=-0.02, ymax=0.17,
	ytick={0.0, 0.02, 0.04, 0.06, 0.08, 0.10, 0.12, 0.14, 0.16},
	xlabel=Number of Qubits,
	ylabel={Overlap for $0.90$-opt, $\braket{\mathbf{O}_{90}}$},
	ymajorgrids=true,
	grid style=dashed,
	legend pos=north east,
	legend style={nodes={scale=0.9, transform shape}},
	width=1\linewidth,
	legend columns=2,
	]
\addplot [color=RoyalBlue, mark=diamond*, mark size=1pt, mark options=solid, line width=1.2pt, error bars/.cd,y dir=both,y explicit] plot coordinates {
	(12, 0.11150742445102253) +- (0.012969898222633286, 0.012969898222633286)
	(14, 0.06597897021069483) +- (0.022457476960037737, 0.022457476960037737)
	(16, 0.03405744375906588) +- (0.018224787116232072, 0.018224787116232072)
	(19, 0.03868712104904616) +- (0.007478788201607757, 0.007478788201607757)
};
\addplot [color=WildStrawberry, mark=square*, mark size=1pt, mark options=solid, line width=1.2pt, error bars/.cd,y dir=both,y explicit] plot coordinates {
	(12, 0.0931280933694546) +- (0.0393482159423643, 0.0393482159423643)
	(14, 0.07710972159220023) +- (0.04224013862354281, 0.04224013862354281)
	(16, 0.03320259048312084) +- (0.016379961434591347, 0.016379961434591347)
	(19, 0.033349770056975904) +- (0.02367097951567167, 0.02367097951567167)
};
\addplot [color=YellowOrange,  mark=triangle*, mark size=1pt, mark options=solid, line width=1.2pt, error bars/.cd,y dir=both,y explicit] plot coordinates {
	(12, 0.09508471247553399) +- (0.0546908703706919, 0.0546908703706919)
	(14, 0.06762522460964612) +- (0.04403218477644605, 0.04403218477644605)
	(16, 0.044698804531462494) +- (0.029619247737470388, 0.029619247737470388)
	(19, 0.02647234813472374) +- (0.016155131833463985, 0.016155131833463985)
};
\addplot [color=BlueGreen, mark=*, mark size=1pt, mark options=solid, line width=1.2pt, error bars/.cd,y dir=both,y explicit] plot coordinates {
	(12, 0.09149506359357551) +- (0.03512972340310523, 0.03512972340310523)
	(14, 0.055034886831774876) +- (0.04695638315344292, 0.04695638315344292)
	(16, 0.03985363938111113) +- (0.025601268561150047, 0.025601268561150047)
	(19, 0.0319127487322923) +- (0.023363101437698063, 0.023363101437698063)
};
\addplot [color=Periwinkle,  mark=otimes*, mark size=1pt, mark options=solid, line width=1.2pt, error bars/.cd,y dir=both,y explicit] plot coordinates {
	(12, 0.09186549528789736) +- (0.06267546109397149, 0.06267546109397149)
	(14, 0.03942069173741514) +- (0.01817021513543379, 0.01817021513543379)
	(16, 0.026665295670559815) +- (0.02159062341004863, 0.02159062341004863)
	(19, 0.019155019124742746) +- (0.016400215278273986, 0.016400215278273986)
};
\addplot [color=Bittersweet,  mark=pentagon, mark size=1pt, mark options=solid, line width=1.2pt, error bars/.cd,y dir=both,y explicit] plot coordinates {
	(12, 0.06988260948622693) +- (0.027841957867996345, 0.027841957867996345)
	(14, 0.04072326078151413) +- (0.024260233604377957, 0.024260233604377957)
	(16, 0.024327180093027718) +- (0.018231554113158795, 0.018231554113158795)
	(19, 0.014566368742592883) +- (0.020640210940985, 0.020640210940985)
};
\legend{p=$1$, p=$2$, p=$3$, p=$4$, p=$5$, p=$6$}
\end{axis}
\end{tikzpicture}
\caption{\small{WS QAOA.}}
\label{res_wsqaoa_overlap}
\end{subfigure}
\hfill
\begin{subfigure}{0.32\linewidth}
\hspace{2.1em}
\begin{tikzpicture}[trim axis left, trim axis right, scale=0.8, every node/.style={scale=0.8}]
\begin{axis}[
	tick label style={/pgf/number format/fixed},scale only axis,
	xmin=11.7, xmax=19.3,
	xtick={12, 14, 16, 19},
	ymin=-0.02, ymax=0.17,
	ytick={0.0, 0.02, 0.04, 0.06, 0.08, 0.10, 0.12, 0.14, 0.16},
	xlabel=Number of Qubits,
	ylabel style={align=center},
	ylabel={Overlap for $0.90$-opt, $\braket{\mathbf{O}_{90}}$},
	ymajorgrids=true,
	grid style=dashed,
 	legend pos=south west,
 	legend style={nodes={scale=0.9, transform shape}},
	width=1\linewidth,
	legend columns=2,
	]
\addplot [color=RoyalBlue, mark=diamond*, mark size=1pt, mark options=solid, line width=1.2pt, error bars/.cd,y dir=both,y explicit] plot coordinates {
(12, 0.12735339157151063) +- (0.014668175776127664, 0.014668175776127664)
(14, 0.09320225341385785) +- (0.029403702799654142, 0.029403702799654142)
(16, 0.057756489343563075) +- (0.027745697046215788, 0.027745697046215788)
(19, 0.0579668103852858) +- (0.012995203362106177, 0.012995203362106177)
};
\addplot [color=WildStrawberry, mark=diamond*, mark size=1pt, mark options=solid, line width=1.2pt, error bars/.cd,y dir=both,y explicit] plot coordinates {
(12, 0.10650755178386824) +- (0.019175365319255226, 0.019175365319255226)
(14, 0.08695549758621843) +- (0.038309402753860224, 0.038309402753860224)
(16, 0.05982939393996509) +- (0.028863480861950012, 0.028863480861950012)
(19, 0.038444963228499135) +- (0.012926053739529628, 0.012926053739529628)
};
\addplot [color=YellowOrange, mark=diamond*, mark size=1pt, mark options=solid, line width=1.2pt, error bars/.cd,y dir=both,y explicit] plot coordinates {
(12, 0.11153924255062989) +- (0.020783463900786003, 0.020783463900786003)
(14, 0.075899388659209) +- (0.05387280748130998, 0.05387280748130998)
(16, 0.06063154046455386) +- (0.015235946079992877, 0.015235946079992877)
(19, 0.04094751394531448) +- (0.023067787777882834, 0.023067787777882834)
};
\addplot [color=BlueGreen, mark=diamond*, mark size=1pt, mark options=solid, line width=1.2pt, error bars/.cd,y dir=both,y explicit] plot coordinates {
(12, 0.08764006134457307) +- (0.026259667334904188, 0.026259667334904188)
(14, 0.08900838322548646) +- (0.05717735042055952, 0.05717735042055952)
(16, 0.06591996928497283) +- (0.04226433216976422, 0.04226433216976422)
(19, 0.03954853206758019) +- (0.01890863113134525, 0.01890863113134525)
};
\addplot [color=Periwinkle, mark=diamond*, mark size=1pt, mark options=solid, line width=1.2pt, error bars/.cd,y dir=both,y explicit] plot coordinates {
(12, 0.10658610186419688) +- (0.05157127756894462, 0.05157127756894462)
(14, 0.12894845718388845) +- (0.05425608278057956, 0.05425608278057956)
(16, 0.05460522771499261) +- (0.051375046292671096, 0.051375046292671096)
(19, 0.03367559375970898) +- (0.020922885862596996, 0.020922885862596996)
};
\addplot [color=Bittersweet, mark=diamond*, mark size=1pt, mark options=solid, line width=1.2pt, error bars/.cd,y dir=both,y explicit] plot coordinates {
(12, 0.10491764289896496) +- (0.03990222382274216, 0.03990222382274216)
(14, 0.1007578017409223) +- (0.05369106579335477, 0.05369106579335477)
(16, 0.06605289538397906) +- (0.03959336526729964, 0.03959336526729964)
(19, 0.0519645172665328) +- (0.036249572555118396, 0.036249572555118396)
};
\legend{p=$1$, p=$2$, p=$3$, p=$4$, p=$5$, p=$6$}
\end{axis}
\end{tikzpicture}
\centering
\caption{\small{WS-Init-QAOA.}}
\label{res_ws_init_qaoa_overlap}
\end{subfigure}%
\caption{\small{Standard QAOA, Warm-start QAOA and WS-Init-QAOA overlaps with 0.90-opt results for different problem sizes and number of QAOA layers.}}
\label{res_overlap_qaoa}
\end{figure}
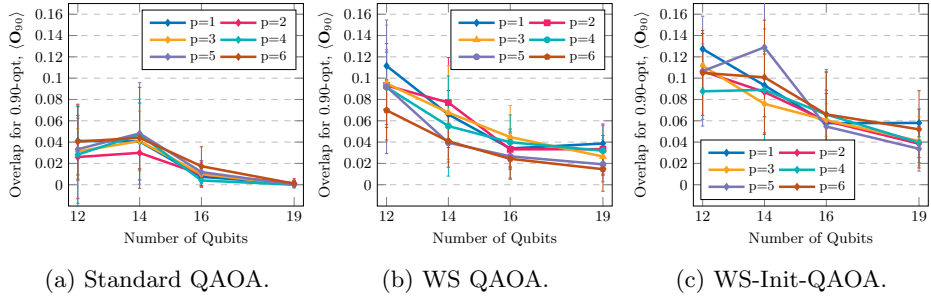
We carried out tests for the three QAOA based algorithms for the four different problem instances with a number of QAOA layers ranging from $p=1$ to $p=6$. For each problem instance, we randomly initialize the $\boldsymbol{\gamma}$ and $\boldsymbol{\beta}$ angles in the range $[-\pi,\pi]$. Additionally, to better assess the convergence of results, we repeat each test $20$ times and report the average and the standard deviation values. 

Figure~\ref{res_overlap_qaoa} presents 0.90-opt overlap $\braket{\mathbf{O}_{90}}$ results for QAOA, WS-QAOA and WS-Init-QAOA against problem instances ranging from $12$ to $19$ qubits, for varying numbers of QAOA layers $p$, along with their standard deviation. The overlap values evidently decrease by increasing the problem sizes, reaching $\approx 0$ for the last scenario with 19 qubits for standard QAOA. Increasing the number of layers seems to present none to minimal improvement in the solution values. This can be seen as an indicator that adding more layers does not increase the subspace of states explored by the QAOA-circuit. The overlap values obtained for WS-QAOA (Figure~\ref{res_wsqaoa_overlap}) are much better than for QAOA. As discussed in Section~\ref{algo_ws_qaoa}, WS-QAOA differs from the standard QAOA in two aspects, the initial state $\ket{\phi^*}$ and the mixing Hamiltonian $\hat{H}_M^{ws}$ whose ground state eigenvector is $\ket{\phi^*}$. Our results show that these two modifications certainly lead to a better performance over standard QAOA.

Remarkably though, WS-Init-QAOA (Figure~\ref{res_ws_init_qaoa_overlap}) seems to perform much better for all the instances, in comparison to WS-QAOA. This is an interesting result considering the mixer Hamiltonian for the WS-Init-QAOA brings the quantum states to their superposition position state (similar to standard QAOA). On the other hand, the mixer Hamiltonian for WS-QAOA is designed to bring the quantum states to the optimum continuous solution. Regardless, it is apparent that just improving the choice of initial state leads to even better results compared to modifying the mixer operator along with the initial value. We would like to emphasize this aspect and suggest that the good solutions obtained by WS-QAOA are mainly due to the classical pre-processing (\emph{i.e.}, the continuous solutions to the relaxed QUBO), and not due to the modified mixing Hamiltonian ($\hat{H}_M^{ws}$).
\begin{figure}[ht!]
\begin{subfigure}{0.32\linewidth}
\hspace{2em}
\begin{tikzpicture}[trim axis left, trim axis right, scale=0.8, every node/.style={scale=0.8}]
\begin{axis}[
	tick label style={/pgf/number format/fixed},
	scale only axis,
	xmin=11.7, xmax=19.3,
	xtick={12, 14, 16, 19},
	ymin=50, ymax=103,
	ytick={50, 60, 70, 80, 90, 100},
	xlabel=Number of Qubits,
	ylabel style={align=center},
	ylabel={Closeness to optimum (\%)},
	ymajorgrids=true,
	grid style=dashed,
	legend pos=south west,
	legend style={nodes={scale=0.9, transform shape}},
	width=1\linewidth,
	legend columns=2,
	]
\addplot [color=RoyalBlue, mark=diamond*, mark size=1pt, mark options=solid, line width=1.2pt, error bars/.cd,y dir=both,y explicit] plot coordinates {
(12, 96.36363636363636) +- (4.4536177141512345, 4.4536177141512345)
(14, 94.16666666666667) +- (9.166666666666664, 9.166666666666664)
(16, 83.84615384615385) +- (9.389658165949001, 9.389658165949001)
(19, 66.92307692307692) +- (22.040844279837543, 22.040844279837543)
};
\addplot [color=WildStrawberry, mark=diamond*, mark size=1pt, mark options=solid, line width=1.2pt, error bars/.cd,y dir=both,y explicit] plot coordinates {
(12, 91.36363636363637) +- (9.192612916434856, 9.192612916434856)
(14, 95.83333333333333) +- (5.5901699437494745, 5.5901699437494745)
(16, 86.92307692307693) +- (12.427303401079625, 12.427303401079625)
(19, 63.84615384615385) +- (20.3664650690365, 20.3664650690365)
};
\addplot [color=YellowOrange, mark=diamond*, mark size=1pt, mark options=solid, line width=1.2pt, error bars/.cd,y dir=both,y explicit] plot coordinates {
(12, 96.36363636363636) +- (4.4536177141512345, 4.4536177141512345)
(14, 86.66666666666666) +- (30.550504633038933, 30.550504633038933)
(16, 89.99999999999999) +- (7.730673554708376, 7.730673554708376)
(19, 55.38461538461538) +- (18.461538461538463, 18.461538461538463)
};
\addplot [color=BlueGreen, mark=diamond*, mark size=1pt, mark options=solid, line width=1.2pt, error bars/.cd,y dir=both,y explicit] plot coordinates {
(12, 92.72727272727272) +- (7.925270806437584, 7.925270806437584)
(14, 97.5) +- (5.335936864527373, 5.335936864527373)
(16, 84.61538461538461) +- (10.320313742306723, 10.320313742306723)
(19, 59.230769230769226) +- (15.783295791294764, 15.783295791294764)
};
\addplot [color=Periwinkle, mark=diamond*, mark size=1pt, mark options=solid, line width=1.2pt, error bars/.cd,y dir=both,y explicit] plot coordinates {
(12, 94.0909090909091) +- (9.968960090664233, 9.968960090664233)
(14, 90.83333333333334) +- (17.26026264767331, 17.26026264767331)
(16, 90.76923076923077) +- (10.204999354939693, 10.204999354939693)
(19, 57.69230769230769) +- (33.39737437860371, 33.39737437860371)
};
\addplot [color=Bittersweet, mark=diamond*, mark size=1pt, mark options=solid, line width=1.2pt, error bars/.cd,y dir=both,y explicit] plot coordinates {
(12, 94.54545454545455) +- (7.27272727272727, 7.27272727272727)
(14, 94.16666666666666) +- (8.374896350934073, 8.374896350934073)
(16, 92.3076923076923) +- (9.101661204768636, 9.101661204768636)
(19, 73.07692307692307) +- (16.22694085363768, 16.22694085363768)
};	
\legend{p=$1$, p=$2$, p=$3$, p=$4$, p=$5$, p=$6$}
\end{axis}
\end{tikzpicture}
\centering
\caption{\small{Standard QAOA.}}
\label{res_qaoa_closeness}
\end{subfigure}%
\hfill
\begin{subfigure}{.32\linewidth}
\hspace{2.1em}
\centering
\begin{tikzpicture}[trim axis left, trim axis right, scale=0.8, every node/.style={scale=0.8}]
\begin{axis}[
tick label style={/pgf/number format/fixed},
scale only axis,
	xmin=11.7, xmax=19.3,
	xtick={12, 14, 16, 19},
	ymin=50, ymax=103,
	ytick={50, 60, 70, 80, 90, 100},
	xlabel=Number of Qubits,
	ylabel={Closeness to optimum (\%)},
	ymajorgrids=true,
	grid style=dashed,
	legend pos=south west,
	legend style={nodes={scale=0.9, transform shape}},
	width=1\linewidth,
	legend columns=2,
	]
\addplot [color=RoyalBlue, mark=diamond*, mark size=1pt, mark options=solid, line width=1.2pt, error bars/.cd,y dir=both,y explicit] plot coordinates {
(12, 100.0) +- (0.0, 0.0)
(14, 100.0) +- (0.0, 0.0)
(16, 100.0) +- (0.0, 0.0)
(19, 100.0) +- (0.0, 0.0)
};
\addplot [color=WildStrawberry, mark=diamond*, mark size=1pt, mark options=solid, line width=1.2pt, error bars/.cd,y dir=both,y explicit] plot coordinates {
(12, 100.0) +- (0.0, 0.0)
(14, 94.99999999999999) +- (5.527707983925667, 5.527707983925667)
(16, 97.6923076923077) +- (4.925480182640654, 4.925480182640654)
(19, 97.6923076923077) +- (6.92307692307692, 6.92307692307692)
};
\addplot [color=YellowOrange, mark=diamond*, mark size=1pt, mark options=solid, line width=1.2pt, error bars/.cd,y dir=both,y explicit] plot coordinates {
(12, 99.0909090909091) +- (2.727272727272728, 2.727272727272728)
(14, 97.5) +- (5.335936864527373, 5.335936864527373)
(16, 97.6923076923077) +- (3.5250582268891084, 3.5250582268891084)
(19, 95.38461538461539) +- (9.230769230769228, 9.230769230769228)
};
\addplot [color=BlueGreen, mark=diamond*, mark size=1pt, mark options=solid, line width=1.2pt, error bars/.cd,y dir=both,y explicit] plot coordinates {
(12, 96.36363636363636) +- (4.4536177141512345, 4.4536177141512345)
(14, 95.83333333333333) +- (5.5901699437494745, 5.5901699437494745)
(16, 95.38461538461539) +- (6.153846153846155, 6.153846153846155)
(19, 88.46153846153845) +- (17.28631158018787, 17.28631158018787)
};
\addplot [color=Periwinkle, mark=diamond*, mark size=1pt, mark options=solid, line width=1.2pt, error bars/.cd,y dir=both,y explicit] plot coordinates {
(12, 98.18181818181817) +- (3.6363636363636376, 3.6363636363636376)
(14, 98.33333333333333) +- (3.3333333333333375, 3.3333333333333375)
(16, 93.84615384615384) +- (5.756395979652218, 5.756395979652218)
(19, 90.0) +- (13.345655056074976, 13.345655056074976)
};
\addplot [color=Bittersweet, mark=diamond*, mark size=1pt, mark options=solid, line width=1.2pt, error bars/.cd,y dir=both,y explicit] plot coordinates {
(12, 99.0909090909091) +- (2.7272727272727284, 2.7272727272727284)
(14, 96.66666666666666) +- (5.527707983925667, 5.527707983925667)
(16, 96.15384615384616) +- (5.160156871153362, 5.160156871153362)
(19, 86.92307692307693) +- (11.94167284327694, 11.94167284327694)
};
\legend{p=$1$, p=$2$, p=$3$, p=$4$, p=$5$, p=$6$}
\end{axis}
\end{tikzpicture}
\subcaption{\small{WS QAOA.}}
\label{res_wsqaoa_closeness}
\end{subfigure}
\hfill
\begin{subfigure}{0.32\linewidth}
\hspace{2.1em}
\begin{tikzpicture}[trim axis left, trim axis right, scale=0.8, every node/.style={scale=0.8}]
\begin{axis}[
tick label style={/pgf/number format/fixed},
scale only axis,
	xmin=11.7, xmax=19.3,
	xtick={12, 14, 16, 19},
	ymin=50, ymax=103,
	ytick={50, 60, 70, 80, 90, 100},
	xlabel=Number of Qubits,
	ylabel={Closeness to optimum (\%)},
	ymajorgrids=true,
	grid style=dashed,
	legend pos=south west,
	legend style={nodes={scale=0.9, transform shape}},
	width=1\linewidth,
	legend columns=2,
	]
\addplot [color=RoyalBlue, mark=diamond*, mark size=1pt, mark options=solid, line width=1.2pt, error bars/.cd,y dir=both,y explicit] plot coordinates {
(12, 100.0) +- (0.0, 0.0)
(14, 99.16666666666666) +- (2.5000000000000027, 2.5000000000000027)
(16, 99.23076923076924) +- (2.3076923076923075, 2.3076923076923075)
(19, 100.0) +- (0.0, 0.0)
};
\addplot [color=WildStrawberry, mark=diamond*, mark size=1pt, mark options=solid, line width=1.2pt, error bars/.cd,y dir=both,y explicit] plot coordinates {
(12, 100.0) +- (0.0, 0.0)
(14, 98.33333333333333) +- (3.3333333333333375, 3.3333333333333375)
(16, 99.23076923076924) +- (2.307692307692308, 2.307692307692308)
(19, 100.0) +- (0.0, 0.0)
};
\addplot [color=YellowOrange, mark=diamond*, mark size=1pt, mark options=solid, line width=1.2pt, error bars/.cd,y dir=both,y explicit] plot coordinates {
(12, 100.0) +- (0.0, 0.0)
(14, 96.66666666666666) +- (4.082482904638635, 4.082482904638635)
(16, 100.0) +- (0.0, 0.0)
(19, 93.84615384615384) +- (18.461538461538463, 18.461538461538463)
};
\addplot [color=BlueGreen, mark=diamond*, mark size=1pt, mark options=solid, line width=1.2pt, error bars/.cd,y dir=both,y explicit] plot coordinates {
(12, 100.0) +- (0.0, 0.0)
(14, 97.5) +- (5.335936864527373, 5.335936864527373)
(16, 96.92307692307693) +- (6.153846153846154, 6.153846153846154)
(19, 100.0) +- (0.0, 0.0)
};
\addplot [color=Periwinkle, mark=diamond*, mark size=1pt, mark options=solid, line width=1.2pt, error bars/.cd,y dir=both,y explicit] plot coordinates {
(12, 100.0) +- (0.0, 0.0)
(14, 98.33333333333333) +- (3.3333333333333375, 3.3333333333333375)
(16, 99.23076923076924) +- (2.3076923076923075, 2.3076923076923075)
(19, 97.6923076923077) +- (6.92307692307692, 6.92307692307692)
};
\addplot [color=Bittersweet, mark=diamond*, mark size=1pt, mark options=solid, line width=1.2pt, error bars/.cd,y dir=both,y explicit] plot coordinates {
(12, 99.0909090909091) +- (2.7272727272727284, 2.7272727272727284)
(14, 100.0) +- (0.0, 0.0)
(16, 99.23076923076924) +- (2.307692307692308, 2.307692307692308)
(19, 100.0) +- (0.0, 0.0)
};
\legend{p=$1$, p=$2$, p=$3$, p=$4$, p=$5$, p=$6$}
\end{axis}
\end{tikzpicture}
\centering
\caption{\small{WS-Init-QAOA.}}
\label{res_ws_init_qaoa_closeness}
\end{subfigure}%
\caption{\small{Closeness to optimum of the best solutions obtained from standard QAOA, warm-start QAOA and WS-Init-QAOA for different problem sizes and number of QAOA layers.}}
\label{res_closeness}
\end{figure}
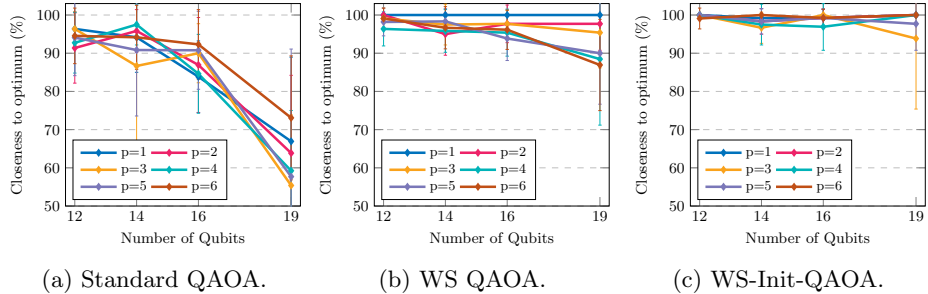
As far as the closeness of the best solution to the optimum is concerned, both warm-start approaches outperform standard QAOA, as shown in Figure~\ref{res_closeness}. While standard QAOA (Figure~\ref{res_qaoa_closeness}) values drop below the $90\%$-mark already for $14$ qubit problems, the WS-QAOA (Figure~\ref{res_wsqaoa_closeness}), maintains $90\%$ of the optimal solution also for $19$ qubits and the WS-Init-QAOA results remain even well above $90\%$ (Figure~\ref{res_ws_init_qaoa_closeness}). The number of layers seems to have only a minor influence on the quality of results. Regarding the stability of the results, we notice that the standard deviation of both WS-QAOA results is smaller than in the standard QAOA. WS-Init-QAOA seems to again provide better values than the WS-QAOA.

In summary, while the number of QAOA layers does not have a considerable influence, using warm-start QAOA increases the quality of the results. Moreover, from the comparison of the two warm-start approaches we see that choosing a better initial state for the quantum circuit leads to significant improvements of the result quality. On the other hand, the additional modification of the mixing Hamiltonian incorporated in the WS-QAOA algorithm does not seem to have a distinct effect, since the WS-Init-QAOA approach, that just uses a different initial state, gives the best results. Other variations of QAOA including better initial rotation angles, as suggested in~\cite{zhou}, need to be looked at to fully understand the true potential of QAOA for multi-knapsack problems. 

\subsection{Results from VQE}
\begin{figure}
    \centering
    \includegraphics[scale=0.35]{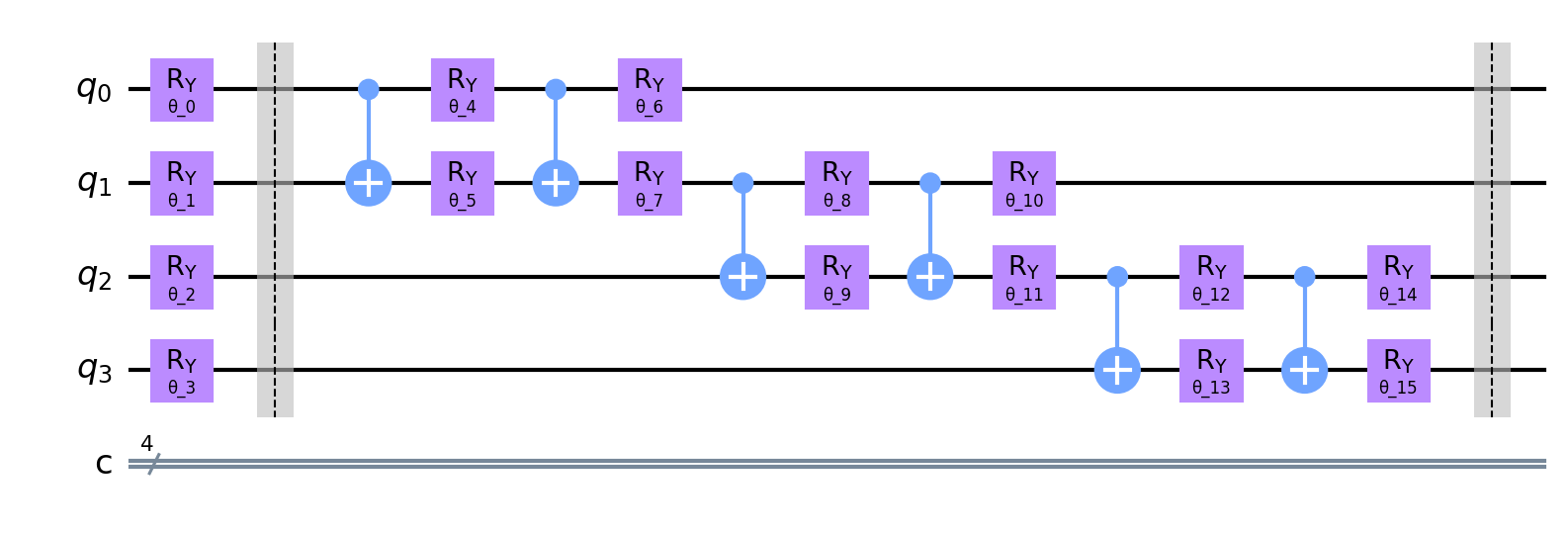}
    \caption{\small{A quantum circuit for $4$-qubit VQE with a single layer.}}
    \label{fig:VQE_circuit.}
\end{figure}

\begin{figure}[hb!]
\begin{subfigure}{0.4\linewidth}
\hspace{2em}
\begin{tikzpicture}[trim axis left, trim axis right, scale=0.8, every node/.style={scale=0.8}]
\begin{axis}[
tick label style={/pgf/number format/fixed},
scale only axis,
xmin=11.8, xmax=19.2,
xtick={12, 14, 16, 19},
ymin=-0.22, ymax=0.5,
ytick={-0.2, -0.1, 0.0, 0.1, 0.2, 0.3, 0.4},
xlabel=Number of Qubits,
ylabel style={align=center},
ylabel={Overlap for $0.90$-opt, $\braket{\mathbf{O}_{90}}$},
ymajorgrids=true,
grid style=dashed,
legend pos=north east,
width=1\linewidth,
legend columns=1,
]
\addplot [color=RoyalBlue, mark=diamond*, mark size=1pt, mark options=solid, line width=1.2pt, error bars/.cd,y dir=both,y explicit] plot coordinates {
(12, 0.06517965826961752)+-(0.1439698716551375, 0.1439698716551375)
(14, 0.1653700338821918)+-(0.2917767494365904, 0.2917767494365904)
(16, 0.012661306192475432)+-(0.04678497046193481, 0.04678497046193481)
(19, 0.0)+-(0.0, 0.0)
};
\addplot [color=YellowOrange, mark=square*, mark size=1pt, mark options=solid, line width=1.2pt, error bars/.cd,y dir=both,y explicit] plot coordinates {
(12, 0.19265794298064534)+-(0.32048670287338477, 0.32048670287338477)
(14, 0.1334676817986245)+-(0.17977942671191885, 0.17977942671191885)
(16, 0.003118033988749895)+-(0.006728882822950659, 0.006728882822950659)
(19, 0.004565198222618523)+-(0.008897132413772465, 0.008897132413772465)
};

\legend{p=$1$, p=$2$}
\end{axis}
\end{tikzpicture}
\centering
\caption{\small{Overlap with 0.90-opt.}}
\label{vqe_overlap}
\end{subfigure}%
\hspace{2em}
\begin{subfigure}{.4\linewidth}
\hspace{2em}
\centering
\begin{tikzpicture}[trim axis left, trim axis right, scale=0.8, every node/.style={scale=0.8}]
\begin{axis}[
tick label style={/pgf/number format/fixed},
scale only axis,
	xmin=11.8, xmax=19.2,
	xtick={12, 14, 16, 19},
	ymin=50, ymax=103,
	ytick={50, 60, 70, 80, 90, 100},
	xlabel=Number of Qubits,
	ylabel={Closeness to optimum (\%)},
	ymajorgrids=true,
	grid style=dashed,
	legend pos=north east,
	width=1\linewidth,
	legend columns=1,
	]
	
\addplot [color=RoyalBlue, mark=diamond*, mark size=1pt, mark options=solid, line width=1.2pt, error bars/.cd,y dir=both,y explicit] plot coordinates {
(12, 86.5909090909091)+-(13.672301129722758, 13.672301129722758)
(14, 86.5)+-(20.802644062714723, 20.802644062714723)
(16, 76.53846153846156)+-(14.488314001413356, 14.488314001413356)
(19, 73.84615384615387)+-(13.18956015344788, 13.18956015344788)
};
\addplot [color=YellowOrange, mark=square*, mark size=1pt, mark options=solid, line width=1.2pt, error bars/.cd,y dir=both,y explicit] plot coordinates {
(12, 91.5909090909091)+-(9.987078428509014, 9.987078428509014)
(14, 100.0)+-(0, 0)
(16, 83.46153846153845)+-(10.679571828396014, 10.679571828396014)
(19, 78.84615384615385)+-(16.295169357165275, 16.295169357165275)
};
\legend{p=$1$, p=$2$}
\end{axis}
\end{tikzpicture}
\caption{\small{Closeness to optimum.}}
\label{vqe_approx}
\end{subfigure}
\captionsetup{width=.85\linewidth}
\centering
\caption{\small{Overlap and closeness results for VQE algorithm for the multi-knapsack problem for $1$ and $2$ layers circuit ansatz.}}
\label{res_vqe}
\end{figure}
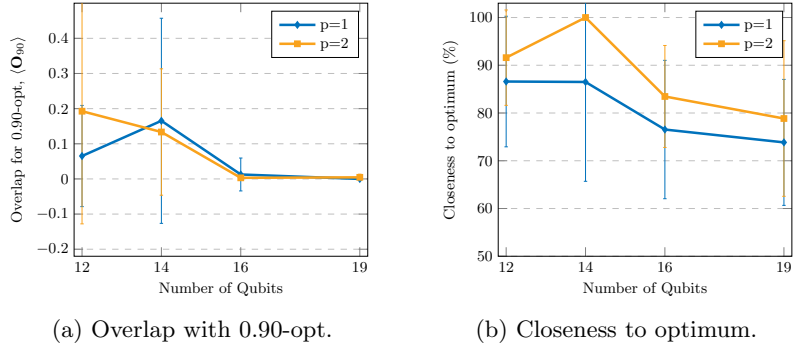

The ansatz circuit chosen for the VQE experiments is adopted from the work of Liu~\emph{et al.}~\cite{Liu_2022} and consists of parameterized single-qubit rotations and two-qubit entangling gates without parameterization, schematically shown in Figure \ref{fig:VQE_circuit.}.
We conducted the experiments with $p=1$ and $p=2$ layers, sampling 10,000 shots from the corresponding quantum circuits. All results are averaged over $20$ runs, and the error bars in Figure~\ref{res_vqe} indicate the standard deviation. Parameters are randomly initialized in the range $[0,2\pi]$, and the COBYLA optimizer is used for the classical parameter tuning. Compared to the QAOA results, the VQE reaches slightly worse approximation rates in general. However, when good solutions are found, they are sampled with higher probability compared to QAOA. The quality of the solutions improves when we take an ansatz circuit with $2$ layers, although this results in almost twice as many parameters and thus a substantial calculation overhead on the parameter tuning side. 

As can be seen from the high number of iterations needed for convergence of the classical optimizer in Table~\ref{tbl:optimizer_iterations}, VQE exhibits a longer runtime than QAOA, even when one accounts for the simulation time (see Section \ref{subsec:runtimes}). These runtimes may serve as a motivation to look for strategies which reduce the number of classical iterations. Overall, the benefits of VQE, \emph{i.e.}, hardware-adapted quantum circuits which suffer less from noise, can only become apparent when running the algorithms on actual quantum hardware. Thus, despite the worse approximation rates and longer run times, VQE deserves further exploration in the context of quantum optimization algorithms.

\subsection{Runtime estimation on quantum devices for QAOA and VQE}\label{subsec:runtimes}
As described in the previous sections, the results for QAOA and VQE are obtained by simulations of the quantum circuits. In order to assess the performance of each quantum algorithm and provide a meaningful comparison with the results from quantum annealing, we estimate the runtimes for standard QAOA and VQE on a quantum processing unit (QPU). A simple model to describe the total runtime $T$ of a variational quantum algorithm is explained in~\cite{Wintersperger2022} and reads:
\begin{equation}
\small
T = n_{\text{iter}}\cdot[n_{\text{samp}}\cdot(t_{\text{circ}}+t_{\text{meas}}) + t_{\text{opt}} + t_{\text{comm}}],\label{eq:runtime}
\end{equation}
where $n_{\text{iter}}$ denotes the number of iterations of the optimizer, $n_{\text{samp}}$ is the number of samples taken to measure the quantum state and $t_{\text{circ}}$, $t_{\text{meas}}$, $t_{\text{opt}}$ describe the times to execute all gates of the quantum circuit, measure all qubits and perform the classical optimization, respectively. 
\begin{wraptable}[10]{r}{0.35\textwidth}
\vspace*{-2em}
\centering
\caption{\small{Gate execution times from Qiskit backend FakeBrooklyn.}}
\vspace*{-1.5em}
\label{tbl:gate_times}
\begin{center}\begin{tabular}{c|c} \hline
\textbf{Gate} & \textbf{Exec. time (ns)}\\ \hline
$R_Z$ & 0 \\
$S_X$ & 35.56  \\
$X$ & 35.56 \\
$CX$ & $370 \pm 80$\\
\hline
\end{tabular}\end{center}
\end{wraptable}
The communication time between the quantum and the classical computer used for optimization is represented by $t_{\text{comm}}$. To provide a rough estimate of the expected runtimes on a QPU, we assume that the times for classical optimization, measurement and communication remain in a similar range as for the simulation and just consider how $t_{\text{circ}}$ is changed. The circuit execution time on a QPU $\tilde{t}_{\text{circ}}$ can be calculated from the gate execution times and the structure of the circuit. As an example, we choose the IBM-Q Brooklyn device consisting of $65$ qubits. The properties such as the topology and gate execution times are estimated using the FakeBrooklyn backend from the FakeProvider module of Qiskit. The corresponding execution times for the native gate set are averaged over all qubits and the mean values are presented in Table~\ref{tbl:gate_times}.
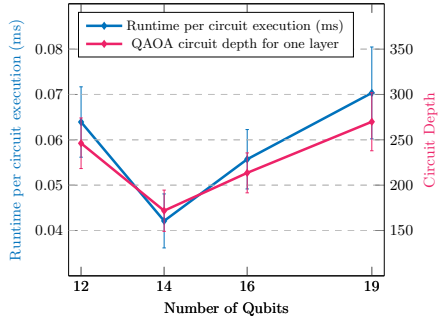
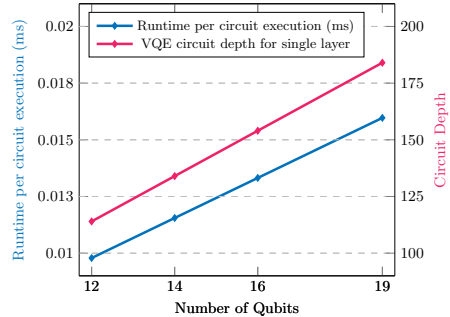
\begin{figure}[hbt!]
\centering
\vspace*{-1.5em}
\begin{subfigure}{0.43\linewidth}
\hspace{2em}
\begin{tikzpicture}[trim axis left, trim axis right, scale=0.8, every node/.style={scale=0.8}]
\begin{axis}[
  axis y line*=left,
  tick label style={/pgf/number format/fixed},scale only axis,
	xmin=11.7, xmax=19.3,
	xtick={12, 14, 16, 19},
	ymin=0.03, ymax=0.09,
	ytick={0.04, 0.05, 0.06, 0.07, 0.08},
	xlabel=Number of Qubits,
	ylabel style={align=center},
	ylabel={\color{RoyalBlue}Runtime per circuit execution (ms)\color{black}},
	ymajorgrids=true,
	grid style=dashed,
	legend pos=north east,
	legend style={nodes={scale=0.9, transform shape}},
	width=1\linewidth,
	legend columns=2,
]
\addplot [color=RoyalBlue, mark=diamond*, mark size=1pt, mark options=solid, line width=1.2pt, error bars/.cd,y dir=both,y explicit] plot coordinates {
(12,0.06392775594999978) +- (0.007761037413910183, 0.007761037413910183)
(14,0.04211533179999993) +- (0.005956781144871354, 0.005956781144871354) 
(16,0.05571538659999983) +- (0.006549948125919716, 0.006549948125919716 )
(19,0.07034099904999971) +- (0.01013160050034065, 0.01013160050034065)
}; \label{plot_one}
\end{axis}
\begin{axis}[
  axis y line*=right,
  tick label style={/pgf/number format/fixed},scale only axis,
	xmin=11.7, xmax=19.3,
	xtick={12, 14, 16, 19},
	ymin=100, ymax=400,
	ytick={150, 200, 250, 300, 350},
	xlabel=Number of Qubits,
	ylabel style={align=center},
	ylabel={\color{WildStrawberry}Circuit Depth\color{black}},
	ymajorgrids=true,
	grid style=dashed,
	legend pos=north west,
	legend style={nodes={scale=0.9, transform shape}},
	width=1\linewidth,
	legend columns=1,
]
\addlegendimage{/pgfplots/refstyle=plot_one}\addlegendentry{Runtime per circuit execution (ms)}%
\addplot [color=WildStrawberry, mark=diamond*, mark size=1pt, mark options=solid, line width=1.2pt, error bars/.cd,y dir=both,y explicit] plot coordinates {
(12, 246.2)  +-  (27.888348821685373, 27.888348821685373)
(14, 171.8)  +-  (22.738953362017348, 22.738953362017348)
(16, 213.55) +-  (22.010168104764674 ,22.010168104764674)
(19, 269.9)  +-  (31.97483385414223, 31.97483385414223)
}; \addlegendentry{QAOA circuit depth for one layer}%
\end{axis}
\end{tikzpicture}
\centering
\caption{\small{QAOA runtime and circuit depth}}
\label{fig:circ_times_and_depths_qaoa}
\end{subfigure}%
\hfill
\begin{subfigure}{.43\linewidth}
\centering
\begin{tikzpicture}[trim axis left, trim axis right, scale=0.8, every node/.style={scale=0.8}]
\begin{axis}[
  axis y line*=left,
  tick label style={/pgf/number format/fixed, /pgf/number format/precision=3},scale only axis,
	xmin=11.7, xmax=19.3,
	xtick={12, 14, 16, 19},
	ymin=0.009, ymax=0.021,
	ytick={0.010, 0.0125, 0.015, 0.0175, 0.0200},
	xlabel=Number of Qubits,
	ylabel style={align=center},
	ylabel={\color{RoyalBlue}Runtime per circuit execution (ms)\color{black}},
	ymajorgrids=true,
	grid style=dashed,
	legend pos=north east,
	legend style={nodes={scale=0.9, transform shape}},
	width=1\linewidth,
	legend columns=2,
]
\addplot [color=RoyalBlue, mark=diamond*, mark size=1pt, mark options=solid, line width=1.2pt, error bars/.cd,y dir=both,y explicit] plot coordinates {
(12, 0.009785893999999996) +- (1.8189894035458564e-18, 1.8189894035458564e-18)
(14, 0.011552217999999994) +- (0, 0) 
(16, 0.013318541999999996) +- (0, 0)
(19, 0.015968027999999995) +- (0, 0)
}; \label{plot_two}
\end{axis}
\begin{axis}[
  axis y line*=right,
  tick label style={/pgf/number format/fixed},scale only axis,
	xmin=11.7, xmax=19.3,
	xtick={12, 14, 16, 19},
	ymin=90, ymax=210,
	ytick={100, 125, 150, 175, 200},
	xlabel=Number of Qubits,
	ylabel style={align=center},
	ylabel={\color{WildStrawberry}Circuit Depth\color{black}},
	ymajorgrids=true,
	grid style=dashed,
	legend pos=north west,
	legend style={nodes={scale=0.9, transform shape}},
	width=1\linewidth,
	legend columns=1,
]
\addlegendimage{/pgfplots/refstyle=plot_two}\addlegendentry{Runtime per circuit execution (ms)}%
\addplot [color=WildStrawberry, mark=diamond*, mark size=1pt, mark options=solid, line width=1.2pt, error bars/.cd,y dir=both,y explicit] plot coordinates {
(12, 114)  +-  (0, 0)
(14, 134)  +-  (0, 0)
(16, 154) +-  (0, 0)
(19, 184)  +-  (0, 0)
}; \addlegendentry{VQE circuit depth for single layer}%
\end{axis}
\end{tikzpicture}
\caption{\small{VQE runtime and circuit depth}}
\label{fig:circ_times_and_depths_vqe}
\end{subfigure}
\caption{\small{Mean circuit execution times $\tilde{t}_{\text{circ}}$ on an IBM-Q device estimated from the gate times and mean circuits depths for QAOA and VQE for $p=1$.}}
\label{fig:circ_times_and_depths_qaoa_vqe}
\end{figure}

The quantum circuits for QAOA and VQE are then transpiled to the FakeBrooklyn backend using the standard Qiskit transpiler with optimization level $3$~\cite{qiskit_2022}. For each circuit, the execution time is derived from a schedule describing the execution of all gates on all qubits, taking into account parallel execution of gates as well as constraints on the timings imposed by two-qubit gates.
The QPU runtimes were estimated for QAOA and VQE with $p=1$, averaging over $20$ compilation runs to account for the stochastic placement of SWAP gates within the standard Qiskit transpilation pass. The mean runtimes $\tilde{t}_{\text{circ}}$ for a single execution of the circuit and their standard deviation are shown in Figure~\ref{fig:circ_times_and_depths_qaoa_vqe}, along with the corresponding mean circuit depths. As described in Section~\ref{sec:vqe}, the VQE algorithm needs less gates, resulting in a mean depth being smaller by about a factor of $2$ compared to QAOA. Due to the different kinds of gates being used, the circuit runtimes are decreased even by a factor of $4$. For VQE, the circuit depth and execution time grow linearly with the problem size, since the structure of the circuit is independent of the specific problem instance and just depends on the number of qubits. In contrast, the circuit depth and execution time of QAOA are larger for the first scenario than for scenario 2 and 3. The reason lies in the different problem structure of scenario 1, which contains only a single knapsack, but more items than the following scenarios which finally leads to a higher number of two-qubit gates in the problem Hamiltonian. 
\begin{figure}[hbt!]
\begin{subfigure}{0.43\linewidth}
\hspace{2em}
\begin{tikzpicture}[trim axis left, trim axis right, scale=0.8, every node/.style={scale=0.8}]
\begin{axis}[
  tick label style={/pgf/number format/fixed},scale only axis,
	xmin=11.7, xmax=19.3,
	xtick={12, 14, 16, 19},
	ymin=-5, ymax=100,
	ytick={0, 5, 10, 20, 40, 60, 80},
	xlabel=Number of Qubits,
	ylabel style={align=center},
	ylabel={Total runtime (seconds)},
	ymajorgrids=true,
	grid style=dashed,
	legend pos=north west,
	legend style={nodes={scale=1, transform shape}},
	width=1\linewidth,
	legend columns=1,
]
\addplot [color=RoyalBlue, mark=diamond*, mark size=1pt, mark options=solid, line width=1.2pt, error bars/.cd,y dir=both,y explicit] plot coordinates {
(12,1.7770517826080323)+-(0.22815635191497183, 0.22815635191497183)
(14,1.8424026012420653)+-(0.27483286375406646, 0.27483286375406646) 
(16,2.2586044788360597)+-(0.15681717495554764, 0.15681717495554764) 
(19,7.496797180175781)+-(1.6578182478343417, 1.6578182478343417)
}; 
\addplot [color=WildStrawberry, mark=diamond*, mark size=1pt, mark options=solid, line width=1.2pt, error bars/.cd,y dir=both,y explicit] plot coordinates {
(12,56.18678959150785) +- (15.402899032322361, 15.402899032322361)
(14,31.87855682964202) +- (6.480175923218559, 6.480175923218559)
(16,36.99922444923596) +- (8.203388885971663, 8.203388885971663)
(19,56.94401865727558) +- (13.007732505285523, 13.007732505285523)
};
\legend{Simulator, Estimation on IBM-Q}
\end{axis}
\end{tikzpicture}
\centering
\caption{\small{Overall runtimes for QAOA}}
 \label{qaoa_scan_prob_C_opt_qiskit}
\end{subfigure}%
\hfill
\begin{subfigure}{.43\linewidth}
\hspace{1em}
\centering
\begin{tikzpicture}[trim axis left, trim axis right, scale=0.8, every node/.style={scale=0.8}]
\begin{axis}[
  tick label style={/pgf/number format/fixed},scale only axis,
	xmin=11.7, xmax=19.3,
	xtick={12, 14, 16, 19},
	ymin=0, ymax=350,
	ytick={50, 100, 150, 200, 250, 300},
	xlabel=Number of Qubits,
	ylabel style={align=center},
	ylabel={Total runtime (seconds)},
	ymajorgrids=true,
	grid style=dashed,
	legend pos=north west,
	legend style={nodes={scale=1, transform shape}},
	width=1\linewidth,
	legend columns=1,
]
\addplot [color=RoyalBlue, mark=diamond*, mark size=1pt, mark options=solid, line width=1.2pt, error bars/.cd,y dir=both,y explicit] 
plot coordinates {
(12,36.86363221) +- (8.83631834011659, 8.83631834011659)
(14,56.644888249999994) +- (8.2613099541776, 8.2613099541776)
(16,77.55096065000001) +- (8.861954609311361, 8.861954609311361)
(19,129.24824123999997) +- (13.898535027506133, 13.898535027506133)
}; 
\addplot [color=WildStrawberry, mark=diamond*, mark size=1pt, mark options=solid, line width=1.2pt, error bars/.cd,y dir=both,y explicit] plot coordinates {
(12,90.68981194199998) +- (14.000162666633736, 14.000162666633736)
(14,136.98870798199997) +- (13.265574565440714, 13.265574565440714)
(16,180.458183218) +- (17.89557667730136, 17.89557667730136)
(19,288.090376076) +- (27.050661936087657, 27.050661936087657)
}; 
\legend{Simulator, Estimation on IBM-Q}
\end{axis}
\end{tikzpicture}
\caption{\small{Overall runtimes for VQE}}
 \label{fig:qaoa_scan_prob_ov90_qiskit}
\end{subfigure}
\caption{\small{Mean overall runtimes on the simulator and estimated on an IBM-Q device for QAOA and VQE with $p=1$. The error bars are computed by error propagation.}}
\label{fig:total_runtimes_qaoa_vqe}
\end{figure}
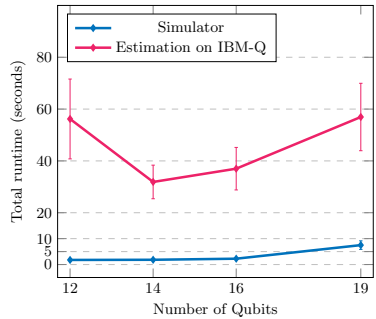
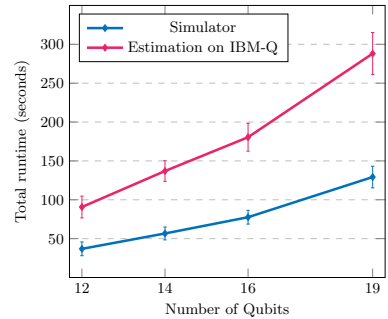

To estimate the overall runtime $\tilde{T}$ on the QPU, we replace $t_{\text{circ}}$ in Equation~\eqref{eq:runtime} with $\tilde{t}_{\text{circ}}$, keeping the other contributions unchanged. The execution times and the overall runtimes on the simulator are obtained from a python profiler, averaging over 5 runs with random initial parameters. 
\begin{wraptable}[10]{l}{0.45\textwidth}
\vspace*{-1.5em}
\centering
\caption{\small{Mean number of optimizer iterations for QAOA and VQE for $p=1$.}}
\vspace*{-1.5em}
\label{tbl:optimizer_iterations}
\begin{center}\begin{tabular}{c|c|c}\hline
\\[-1em]
\textbf{Num. qubits} & $\boldsymbol{n_{\text{iter}}^{\text{qaoa}}}$ & $\boldsymbol{n_{\text{iter}}^{\text{vqe}}}$\\[2pt] \hline
12 & $80 \pm 20$ & $700 \pm 100$ \\
14 & $70 \pm 10$ & $900 \pm 100$ \\
16 & $60 \pm 10$ & $1000 \pm 100$ \\
19 & $80 \pm 20$ & $1400 \pm 200$ \\ 
\hline
\end{tabular}\end{center}
 	\vspace*{1.5em}
\end{wraptable}
In Figure~\ref{fig:total_runtimes_qaoa_vqe}, the resulting runtimes for the simulator are compared with the estimate on the QPU. For QAOA, the runtimes on the QPU are about $30$ times larger than on the simulator. Both runtimes show a similar dependence on the problem size and complexity, following the trend observed in Figure~\ref{fig:circ_times_and_depths_qaoa} for the circuit depth and circuit execution time. In general, we observed no considerable difference in the simulator runtimes and number of optimizer iterations $n_{\text{iter}}$ between standard QAOA, WS-QAOA and WS-init-QAOA.

The overall runtimes for VQE on the simulator and QPU differ only by a factor of $2$-$3$ and are both larger than for QAOA. While the QPU circuit runtimes $\tilde{t}_{\text{circ}}$ are smaller for VQE, the mean circuit runtimes on the simulator lie in the same range as for QAOA with $t_{\text{circ}}^{\text{vqe}} \sim (2.5 - 4.2) \mu$s and $t_{\text{circ}}^{\text{qaoa}} \sim (0.8 - 7.1) \mu$s. However, the VQE algorithm contains more parameters to optimize and thus it takes more iterations to find the optimal parameter values, as shown by the values in Table~\ref{tbl:optimizer_iterations}. This increases the runtimes on the simulator as well as the QPU estimate accordingly.

\subsection{Results from Annealing}
We tested both simulated and quantum annealing, along with the IHS with simulated annealing. For quantum annealing, we focused on two of the devices available on Amazon Braket: \emph{D-Wave 2000Q} (2,048 qubits) and the larger \emph{D-Wave Advantage 6.1} (5,760 qubits). For the IHS, we used simulated annealing with 50 iterations in a single run and set the number of optimization parameters to 12. All annealing algorithms were executed with 1,000 reads and repeated ten times to evaluate the algorithms' stability.

\begin{figure}[ht!]
\centering
\begin{subfigure}{0.4\linewidth}
\centering
\hspace{2em}
\begin{tikzpicture}[trim axis left, trim axis right, scale=0.8, every node/.style = {scale = 0.8} ]
\begin{axis}[
tick label style={/pgf/number format/fixed},
scale only axis,
xmin=11.7, xmax=19.3,
xtick={12, 14, 16, 19},
ymin=0, ymax=0.6,
ytick={0, 0.2, 0.4 , 0.6},
xlabel=Number of logical qubits,
ylabel={Overlap for $0.90$-opt, $\braket{\mathbf{O}_{90}}$},
ymajorgrids=true,
grid style=dashed,
legend pos=north east,
width=1\linewidth,
]

\addplot [color=RoyalBlue,  mark=diamond*, mark size=1pt, mark options=solid, line width=1.2pt, error bars/.cd,y dir=both,y explicit] plot coordinates  {
(12, 0.09767979088731675) +- (0.024914486018514287,0.024914486018514287) 
(14, 0.11882052806752921) +- (0.02947092936274479,0.02947092936274479) 
(16, 0.058635169002042076) +- (0.027119426331649012,0.027119426331649012) 
(19, 0.023445801916009854) +- (0.01667647619703437,0.01667647619703437) 
};
\addplot [color=OrangeRed,  mark=square*, mark size=1pt, mark options=solid, line width=1.2pt, error bars/.cd,y dir=both,y explicit] plot coordinates  {
(12, 0.05935688748494885) +- (0.0329437215076986,0.0329437215076986) 
(14, 0.13441881029392516) +- (0.047511208492379976,0.047511208492379976) 
(16, 0.07074024678938495) +- (0.040767872483896385,0.040767872483896385) 
(19, 0.01943619451055638) +- (0.02629394402071021,0.02629394402071021) 
};
\addplot [color=YellowOrange,  mark=triangle*, mark size=1pt, mark options=solid, line width=1.2pt, error bars/.cd,y dir=both,y explicit] plot coordinates  {
(12, 0.3163527128506354)  +- (0.015668832192301516,0.015668832192301516) 
(14, 0.4337113912089169)  +- (0.010243108014977282,0.010243108014977282) 
(16, 0.37867457953776174) +- (0.010830143730418507,0.010830143730418507) 
(19, 0.2524260956960322)  +- (0.014183958531367174,0.014183958531367174) 
};
\end{axis}
\end{tikzpicture}
\caption{\small{Overlap with 0.90-opt.}}
\centering
\label{fig_anneal_maxval}
\end{subfigure}%
\hspace{2em}
\begin{subfigure}{.4\linewidth}
\hspace{1em}
\centering
\begin{tikzpicture}[trim axis left, trim axis right, scale=0.8, every node/.style={scale=0.8}]

\begin{axis}[
	tick label style={/pgf/number format/fixed},
	scale only axis,
	xmin=11.7, xmax=19.3,
	xtick={12, 14, 16, 19},
	ymin=70, ymax=105,
	ytick={75, 85, 95, 100},
	xlabel=Number of logical qubits,
	ylabel={Closeness to optimum (\%)},
	ymajorgrids=true,
	grid style=dashed,
	width=1\linewidth
	]

\addplot [color=RoyalBlue,  mark=diamond*, mark size=1pt, mark options=solid, line width=1.2pt, error bars/.cd,y dir=both,y explicit] plot coordinates  {
(12, 98.18181818181819) +- (3.8330638305071267,3.8330638305071267) 
(14, 100.0) +- (0.0,0.0) 
(16, 96.15384615384615) +- (5.439282932204212,5.439282932204212) 
(19, 95.38461538461537) +- (7.4314752544561244,7.4314752544561244) 

};
\addplot [color=OrangeRed,  mark=square*, mark size=1pt, mark options=solid, line width=1.2pt, error bars/.cd,y dir=both,y explicit] plot coordinates  {
(12, 94.54545454545456) +- (6.356417261637281,6.356417261637281) 
(14, 100.0) +- (0.0,0.0) 
(16, 94.61538461538461) +- (3.715737627228059,3.715737627228059) 
(19, 88.46153846153847) +- (11.028622137234139,11.028622137234139) 
};
\addplot [color=YellowOrange,  mark=triangle*, mark size=1pt, mark options=solid, line width=1.2pt, error bars/.cd,y dir=both,y explicit] plot coordinates  {
(12, 100.0) +- (0.0,0.0) 
(14, 100.0) +- (0.0,0.0) 
(16, 100.0) +- (0.0,0.0) 
(19, 100.0) +- (0.0,0.0) 

};
\addplot [color=ForestGreen,  mark=*, mark size=1pt, mark options=solid, line width=1.2pt, error bars/.cd,y dir=both,y explicit] plot coordinates  {
(12, 100.0) +- (0.0,0.0) 
(14, 100.0) +- (0.0,0.0) 
(16, 100.0) +- (0.0,0.0) 
(19, 100.0) +- (0.0,0.0) 

};
\end{axis}
\end{tikzpicture}
\caption{\small{Closeness to optimum (\%)}}
\centering
\label{fig_anneal_quality}
\end{subfigure}%
\vspace{2em}
\begin{subfigure}{.4\linewidth}
\hspace{2em}
\centering
\begin{tikzpicture}[trim axis left, trim axis right, scale=0.8, every node/.style={scale=0.8}]

\begin{axis}[
	tick label style={/pgf/number format/fixed},
	scale only axis,
	xtick={12, 14, 16, 19},
	ymin=0, ymax=81,
	ytick={0, 25, 50, 75},
	xlabel=Number of logical qubits,
	ylabel={Number of physical qubits},
	ymajorgrids=true,
	grid style=dashed,
	legend pos=south west,
	width=1\linewidth,
	legend style={font=\footnotesize, at={(0.3,-0.20)},anchor=north west,legend columns=5},
	]

\addlegendentry{-}
\addplot [color=OrangeRed,  mark=square*, mark size=1pt, mark options=solid, line width=1.2pt, error bars/.cd,y dir=both,y explicit] plot coordinates  {
(12, 49.0) +- (1.2472191289246473,1.2472191289246473) 
(14, 36.4) +- (0.699205898780101,0.699205898780101) 
(16, 45.9) +- (1.4491376746189437,1.4491376746189437) 
(19, 67.9) +- (3.381321240777536,3.381321240777536) 

};
\addplot [color=RoyalBlue,  mark=diamond*, mark size=1pt, mark options=solid, line width=1.2pt, error bars/.cd,y dir=both,y explicit] plot coordinates  {
(12, 23.8) +- (0.7888106377466154,0.7888106377466154) 
(14, 20.3) +- (0.6749485577105525,0.6749485577105525) 
(16, 24.9) +- (1.1972189997378642,1.1972189997378642) 
(19, 33.7) +- (1.7669811040931436,1.7669811040931436) 

};
\end{axis}
\end{tikzpicture}
\caption{\small{Number of physical qubits}}
\centering
\label{fig_anneal_qubits}
\end{subfigure}%
\hspace{2em}
\begin{subfigure}{.4\linewidth}
\hspace{1em}
\centering
\begin{tikzpicture}[trim axis left, trim axis right, scale=0.8, every node/.style={scale=0.8}]

\begin{axis}[
tick label style={/pgf/number format/fixed},
scale only axis,
xmin=11.7, xmax=19.3,
xtick={12, 14, 16, 19},
ymin=0, ymax=16,
ytick={0, 5, 10 , 15},
xlabel=Number of logical qubits,
ylabel style={align=center},
ylabel={Runtime (s)},
ymajorgrids=true,
grid style=dashed,
legend pos=south west,
width=1\linewidth,
legend style={font=\small, at={(-1.55,-0.195)},anchor=north west,legend columns=5},
]
\addlegendentry{D-Wave Advantage 6.1}
\addplot [color=RoyalBlue,  mark=diamond*, mark size=1pt, mark options=solid, line width=1.2pt, error bars/.cd,y dir=both,y explicit] plot coordinates  {
(12, 8.4298517) +- (1.6666849807188495,1.6666849807188495) 
(14, 9.1343699) +- (3.1209972651303284,3.1209972651303284) 
(16, 9.9039276) +- (1.6613983490112043,1.6613983490112043) 
(19, 10.2375479) +- (1.6765496661863772,1.6765496661863772) 

};
\addlegendentry{D-Wave 2000Q}
\addplot [color=OrangeRed,  mark=square*, mark size=1pt, mark options=solid, line width=1.2pt, error bars/.cd,y dir=both,y explicit] plot coordinates  {
(12, 6.816035100000001) +- (0.7338664789491414,0.7338664789491414) 
(14, 7.195979599999999) +- (0.3592462050953046,0.3592462050953046) 
(16, 8.5731146) +- (1.2537898745474945,1.2537898745474945) 
(19, 9.6107755) +- (2.1824953830615423,2.1824953830615423) 

};
\addlegendentry{Simulated Annealing}
\addplot [color=YellowOrange,  mark=triangle*, mark size=1pt, mark options=solid, line width=1.2pt, error bars/.cd,y dir=both,y explicit] plot coordinates  {
(12, 0.3353367) +- (0.006878461117462583,0.006878461117462583) 
(14, 0.3783162) +- (0.005662188873964262,0.005662188873964262) 
(16, 0.4247886) +- (0.004772052113900051,0.004772052113900051) 
(19, 0.6124164999999999) +- (0.012769272260217325,0.012769272260217325) 

};
\addlegendentry{IHS with Sim. Anneal.}
\addplot [color=ForestGreen,  mark=*, mark size=1pt, mark options=solid, line width=1.2pt, error bars/.cd,y dir=both,y explicit] plot coordinates  {
(12, 11.221363499999999) +- (0.024831494478898625,0.024831494478898625) 
(14, 10.8030916) +- (0.34651364094976994,0.34651364094976994) 
(16, 11.150483900000001) +- (1.125870072066385,1.125870072066385) 
(19, 11.9443742) +- (3.0995021366997997,3.0995021366997997) 

};

\end{axis}
\end{tikzpicture}
\caption{\small{Runtime (seconds)}}
\centering
\label{fig_anneal_runtime}
\end{subfigure}
\captionsetup{width=.85\linewidth}
\centering
\caption{\small{Overlap, closeness to optimum, number of required physical qubits for annealing and runtime.}}
\end{figure}
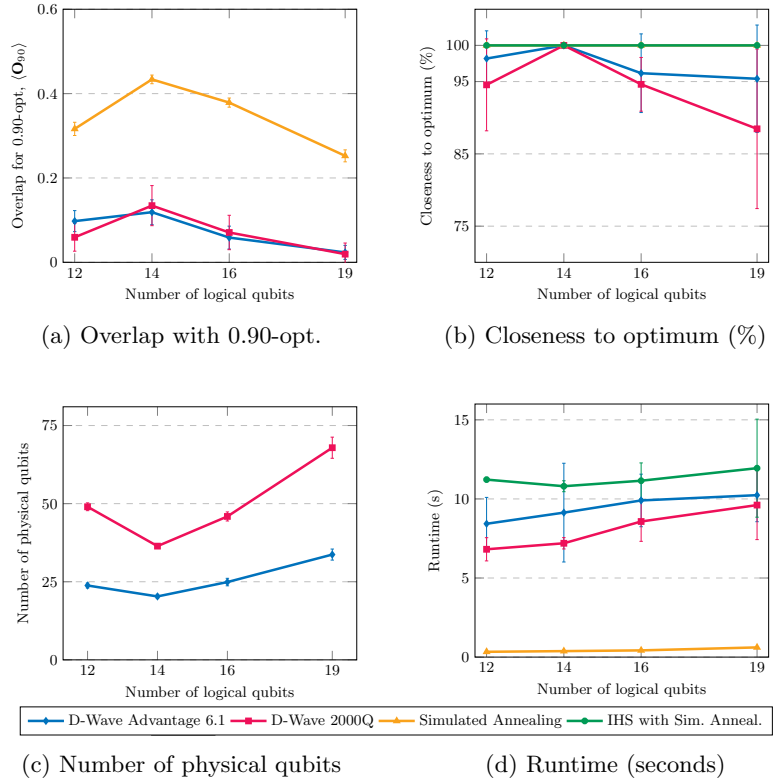

The $\braket{\mathbf{O}_{90}}$ results displayed in Figure~\ref{fig_anneal_maxval} show a higher overlap for simulated annealing - between $0.25\pm0.01$ in scenarios with 19 qubits and $0.43\pm0.01$ with 14 qubits - compared to the quantum annealing options which perform similar and do not exceed $0.14\pm0.05$ with \textit{D-Wave 2000Q} in scenarios with 14 qubits. For IHS, the 0.90-opt overlap cannot be computed in a meaningful way due to the nature of the algorithm.

Figure~\ref{fig_anneal_quality} displays the closeness of the found solution relative to the optimum. The simulated annealing and IHS approaches find optimal solutions for each of the tested knapsack instances. With quantum annealing, the optimal solutions were obtained only for Scenario 2 with 14 qubits. For the other scenarios, \textit{D-Wave Advantage 6.1} exhibits solution qualities of $95.4\pm7.4\%$ for the largest problem instance up to $98.2\pm3.8\%$ for Scenario 1 and outperforms \textit{D-Wave 2000Q} yielding $88.5\pm11.0\%$ to $94.5\pm6.4\%$.
Figure~\ref{fig_anneal_qubits} displays the number of physical qubits needed to embed the theoretical qubits from the QUBO, compensating for the limited connectivity of the physical qubits. As expected, the \emph{D-Wave 2000Q} requires more physical qubits to embed the same QUBO than the \emph{D-Wave Advantage 6.1}, as its architecture has lower connectivity\cite{dwave_architecture}. Since finding the embedding is done by a non-deterministic algorithm, the required number physical qubits on each device varies between runs of the same scenario. In scenarios with two knapsacks (Scenarios~2-4), the number of physical qubits grows proportionally to the number of logical qubits needed to model the respective problem. However, in Scenario~1 more physical qubits are required to embed the QUBO than in Scenario~2, even though less logical qubits are necessary to represent the problem with one knapsack and eight items (see Figure~\ref{tab:scenarios}). This shows the importance of considering the physical properties of the particular quantum hardware and not solely looking at the theoretical number of logical qubits.

The runtimes for various annealing approaches are compared in Figure~\ref{fig_anneal_runtime}. For simulated annealing, the average runtimes are below $1s$ for all scenarios, while the runtimes of the IHS are the longest of all annealing approaches considered in this work, ranging from $10.8\pm0.3s$ to $11.9\pm 3.1s$. With runtimes between $6.8\pm0.7s$ and $9.6\pm2.2s$ (\textit{D-Wave Advantage 6.1}) and between $8.4\pm1.7s$ and $10.2\pm1.7s$ (\textit{D-Wave 2000Q}), respectively, quantum annealing approaches are about a factor $3$-$7$ faster than the QPU estimates for QAOA and about $10$-$30$ times faster than the estimates for VQE presented in the previous section. The quantum annealing runtimes grow only moderately with the problem size compared to the pronounced increase observed especially with VQE. Note that in our analysis of quantum annealing runtimes, we also include the communication overhead between the user and the QPU as well as measurement times, which were not included in the estimation for QAOA and VQE. Thus, the runtimes for the gate-based approaches are expected to increase even more in practice, also when adding more layers to the circuits. In general, using a different hardware platform will also influence the runtime, since the gate times depend on the physical realization of the QPU. The circuit runtime also has to be compared to the coherence time of the qubits: while the gate execution times on other platforms such as trapped ions or cold atoms are in general longer than for superconducting QPUs such as the IBM-Q device considered here, those platforms usually also feature longer coherence times. Moreover, realizations based on ions or atoms exhibit better connectivity between the qubits which in turn reduces the amount of gates needed to run the circuit on the QPU and thus also reduces the amount of noise introduced on NISQ-devices~\cite{Wintersperger2022}. Eventually, the absolute values of the runtimes have to be judged along with the quality of the delivered results within the specific context of the application.
\subsection{Comparison of the quantum algorithms}\label{subsec:comparison}
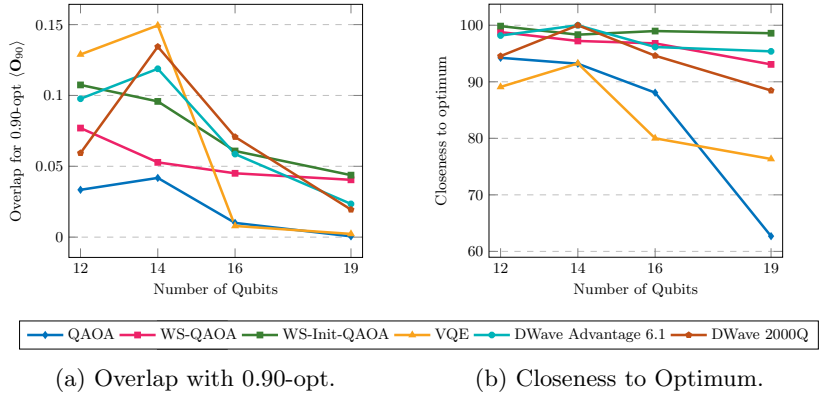
\begin{figure}[hb!]
\centering
\begin{subfigure}{0.4\linewidth}
\hspace{2em}
\begin{tikzpicture}[trim axis left, trim axis right, scale=0.8, every node/.style={scale=0.8}]
\begin{axis}[
tick label style={/pgf/number format/fixed},
scale only axis,
xmin=11.7, xmax=19.3,
xtick={12, 14, 16, 19},
xlabel=Number of Qubits,
ylabel={Overlap for 0.90-opt $\braket{\mathbf{O}_{90}}$},
ymajorgrids=true,
grid style=dashed,
legend pos=north east,
width=1\linewidth,
legend style={font=\small, at={(0.3,-0.25)},anchor=north west,legend columns=5},
]
\addlegendentry{X}
\addplot [color=RoyalBlue, mark=diamond*, mark size=1pt, mark options=solid, line width=1.2pt, error bars/.cd,y dir=both,y explicit] plot coordinates {
(12.0, 0.033403892059939606)
(14.0, 0.04179667549246294)
(16.0, 0.010046338287938064)
(19.0, 0.0005690355937288484)
};

\addplot [color=WildStrawberry, mark=square*, mark size=1pt, mark options=solid, line width=1.2pt, error bars/.cd,y dir=both,y explicit] plot coordinates {
(12.0, 0.07698365399176367) 
(14.0, 0.052774425167078566)
(16.0, 0.04505420870770813)
(19.0, 0.040458312653460275)
};
\addplot [color=OliveGreen, mark=square*, mark size=1pt, mark options=solid, line width=1.2pt, error bars/.cd,y dir=both,y explicit] plot coordinates {
(12.0, 0.10742399866895729)
(14.0, 0.09579529696826376)
(16.0, 0.06079925268867109)
(19.0, 0.04375798844215356)
};
\addplot [color=YellowOrange,  mark=triangle*, mark size=1pt, mark options=solid, line width=1.2pt, error bars/.cd,y dir=both,y explicit] plot coordinates {
(12, 0.12891880062513145)
(14, 0.14941885784040815)
(16, 0.007889670090612663)
(19, 0.0022825991113092615)
};
\addplot [color=BlueGreen, mark=*, mark size=1pt,  line width=1.2pt, error bars/.cd,y dir=both,y explicit] plot coordinates {
(12, 0.09767979088731675)
(14, 0.11882052806752921)
(16, 0.058635169002042076)
(19, 0.023445801916009854)
};
\addplot [color=Bittersweet,  mark=pentagon, mark size=1pt, mark options=solid, line width=1.2pt, error bars/.cd,y dir=both,y explicit] plot coordinates {
(12, 0.05935688748494885)
(14, 0.13441881029392516)
(16, 0.07074024678938495)
(19, 0.01943619451055638)
};
\end{axis}
\end{tikzpicture}
\caption{\small{Overlap with 0.90-opt.}}
\centering
\label{vqe_overlap_all}
\end{subfigure}%
\hspace{2em}
\begin{subfigure}{.4\linewidth}
\hspace{1em}
\centering
\begin{tikzpicture}[trim axis left, trim axis right, scale=0.8, every node/.style={scale=0.8}]
\begin{axis}[
tick label style={/pgf/number format/fixed},
scale only axis,
xmin=11.7, xmax=19.3,
xtick={12, 14, 16, 19},
xlabel=Number of Qubits,
ylabel={Closeness to optimum},
ymajorgrids=true,
grid style=dashed,
legend pos=south west,
width=1\linewidth,
legend style={font=\small, at={(-1.6,-0.25)},anchor=north west,legend columns=6},
]
\addlegendentry{QAOA}
\addplot [color=RoyalBlue, mark=diamond*, mark size=1pt, mark options=solid, line width=1.2pt, error bars/.cd,y dir=both,y explicit] plot coordinates {
(12.0, 94.24242424242425)
(14.0, 93.19444444444444)
(16.0, 88.07692307692308)
(19.0, 62.6923076923077)
};
\addlegendentry{WS-QAOA}
\addplot [color=WildStrawberry, mark=square*, mark size=1pt, mark options=solid, line width=1.2pt, error bars/.cd,y dir=both,y explicit] plot coordinates {
(12.0, 98.7878787878788)
(14.0, 97.22222222222221)
(16.0, 96.79487179487178)
(19.0, 93.07692307692308)
};
\addlegendentry{WS-Init-QAOA}
\addplot [color=OliveGreen, mark=square*, mark size=1pt, mark options=solid, line width=1.2pt, error bars/.cd,y dir=both,y explicit] plot coordinates {
(12.0, 99.84848484848486)
(14.0, 98.33333333333333)
(16.0, 98.97435897435896)
(19.0, 98.58974358974358)
};
\addlegendentry{VQE}
\addplot [color=YellowOrange,  mark=triangle*, mark size=1pt, mark options=solid, line width=1.2pt, error bars/.cd,y dir=both,y explicit] plot coordinates {
(12, 89.0909090909091)
(14, 93.25)
(16, 80.0)
(19, 76.34615384615387)
};
\addlegendentry{DWave Advantage 6.1}
\addplot [color=BlueGreen, mark=*, mark size=1pt,  line width=1.2pt, error bars/.cd,y dir=both,y explicit] plot coordinates {
(12, 98.18181818181819) 
(14, 100.0) 
(16, 96.15384615384615)  
(19, 95.38461538461537)
};
\addlegendentry{DWave 2000Q}
\addplot [color=Bittersweet,  mark=pentagon, mark size=1pt, mark options=solid, line width=1.2pt, error bars/.cd,y dir=both,y explicit] plot coordinates {
(12, 94.54545454545456) 
(14, 100.0) 
(16, 94.61538461538461) 
(19, 88.46153846153847) 
};
\end{axis}
\end{tikzpicture}
\captionsetup{width=.85\linewidth}
\centering
\caption{\small{Closeness to Optimum.}}
\label{res_closeness_all}
\end{subfigure}%
\caption{\small{Average overlap and closeness to optimum values for all the quantum approaches, over different problem sizes.}}
\label{aggregated_results}
\end{figure}

The result quality provided by the different quantum solvers is summarized in Figure~\ref{aggregated_results}, where the results for the gate-based approaches are averaged over different numbers of layers. While the result quality is decreasing with the problem size for all approaches, this effect is most pronounced for VQE and standard QAOA. Overall, the best results are delivered by WS-Init-QAOA and QA on the \textit{D-Wave Advantage 6.1} device, which show quite comparable values for the 0.90-opt overlap and the closeness to optimum. As discussed in the previous section however, the runtimes for QA on a QPU are shorter than the corresponding estimates for QAOA. 

When comparing empirical results from different quantum algorithms, it is important to consider how the differences in implementation can affect both results and their interpretation. For quantum annealers, it is known that noise, embedding overhead, and high precision requirements are all detrimental to performance. Simulated QAOA and VQE do not suffer from these issues, but rather are limited by the quality of parameter settings with classical optimization and statistical sampling accuracy. Conversely, quantum annealing samples are known to primarily populate local minima due to classical effects after the freeze-out point~\cite{dwave_controls}, which is not the case for QAOA and VQE.

\section{Conclusion and outlook}~\label{conclusion}
The aim of this work was to compare different approaches for solving the multi-knapsack problem in view of practical applications. We have defined appropriate measures for this comparison such as the 0.90-opt overlap and have also discussed the implementation of the considered quantum algorithms on real hardware as well as their limitations. The direct comparison of the most common gate-based algorithms with quantum annealing and simulated annealing provided in this work will help to better estimate the potential of these approaches for solving more complex optimization problems such as the knapsack problem in the future. Since practitioners might not have access to various types of quantum computing hardware, the suggested estimation of runtimes on real hardware derived from simulations of quantum circuits can be useful to carry out benchmarks of quantum algorithms.  

Our results show that adapting a standard algorithm such as QAOA can considerably improve the quality of the delivered results. Comparable advantages might be achieved for VQE by finding similar initialization and warm-starting strategies as for QAOA. Deriving optimized starting angles for QAOA (or VQE) as described in~\cite{zhou} or tailoring the schedule of quantum annealing to minimize transfer to higher energy states~\cite{zhou} constitute other promising approaches. Moreover, we can conclude that variational gate-based approaches will profit from better optimization strategies for the circuit parameters to lower the number of iterations, especially in the case of VQE.

As a future work, it would be of great interest to study the QAOA and VQE algorithms incorporating the ideas from Koretsky~\emph{et al.}~\cite{koretsky} and Braine~\emph{et al.}~\cite{braine}, which provide an alternative and qubit-efficient way of formulating inequality constraints in a QUBO model without requiring binary slack bits. These techniques seem to be promising since the convergence of slack bits in all quantum algorithms based on QUBOs is generally hard to achieve. Moreover, it would also be interesting to study and implement the qubit-efficient encoding of optimization problems especially with VQE~\cite{Tan_2021}. 

Tackling realistic problems with millions of qubits, as described in Section~\ref{sec:business_motivation}, is out of scope for the currently available NISQ-devices. Considering the roadmaps of various quantum hardware vendors, however, quantum computers with up to 10,000 qubits might become available within 2-3 years. Thus, using strategies to find optimized problem formulations and algorithms with reduced number of qubits and quantum operations will be of key importance to fit realistic problems onto those intermediate-sized quantum computers. In addition, matching the design of algorithms and quantum computing hardware is seen as another quite promising approach in the NISQ-era. 
\bibliographystyle{splncs03}
\bibliography{literature}
\end{document}